\documentclass[preprint2]{aastex}

\begin{document}

%% LaTeX will automatically break titles if they run longer than
%% one line. However, you may use \\ to force a line break if
%% you desire.

\title{Gamma rays from Nebulae around Recurrent Novae}

%% Use \author, \affil, and the \and command to format
%% author and affiliation information.
%% Note that \email has replaced the old \authoremail command
%% from AASTeX v4.0. You can use \email to mark an email address
%% anywhere in the paper, not just in the front matter.
%% As in the title, use \\ to force line breaks.

\author{W. Bednarek \& J. Sitarek}
\affil{University of Lodz, Faculty of Physics and Applied Informatics, Department of Astrophysics, 90-236 Lodz, ul. Pomorska 149/153, Poland}

\email{wlodzimierz.bednarek@uni.lodz.pl; julian.sitarek@uni.lodz.pl}

%% Notice that each of these authors has alternate affiliations, which
%% are identified by the \altaffilmark after each name.  Specify alternate
%% affiliation information with \altaffiltext, with one command per each
%% affiliation.

%% Mark off your abstract in the ``abstract'' environment. In the manuscript
%% style, abstract will output a Received/Accepted line after the
%% title and affiliation information. No date will appear since the author
%% does not have this information. The dates will be filled in by the
%% editorial office after submission.

\begin{abstract}
Novae were discovered to emit transient $\gamma$ rays during the period of several days to a few weeks after 
initial explosion, indicating presence of acceleration processes of particles in their expanding shells. 
In the case of recurrent novae, electrons can be in principle accelerated in the nova shells for the whole recurrence period of nova producing delayed $\gamma$ ray emission as considered in Bednarek (2022). Here we extend the ideas presented in this article by considering the fate of electrons which diffuse out of 
the shells of novae supplying fresh relativistic electrons to the recurrent nova super-remnants during the 
whole active period of nova ($\ge 10^4$ yrs).
We develop a model for the acceleration of electrons and their escape from the nova shells. The electrons within the recurrent nova super-remnants produce $\gamma$ rays in the comptonization process of the radiation from the red giant companion and the Cosmic Microwave Background Radiation.  
As an example, the case of a symbiotic nova RS Oph (with the recurrence period estimated on $\sim$10-50 yrs) is considered in more detail. Predicted $\gamma$-ray emission from the nova super-remnant around RS Oph is discussed in the context of its observability by satellite experiments 
(i.e. \textit{Fermi}-LAT) as well as current and future Cherenkov telescopes. 
\end{abstract}

%% Keywords should appear after the \end{abstract} command. The uncommented
%% example has been keyed in ApJ style. See the instructions to authors
%% for the journal to which you are submitting your paper to determine
%% what keyword punctuation is appropriate.

%% From the front matter, we move on to the body of the paper.
%% In the first two sections, notice the use of the natbib \citep
%% and \citet commands to identify citations.  The citations are
%% tied to the reference list via symbolic KEYs. The KEY corresponds
%% to the KEY in the \bibitem in the reference list below. We have
%% chosen the first three characters of the first author's name plus
%% the last two numeral of the year of publication as our KEY for
%% each reference.

%% Authors who wish to have the most important objects in their paper
%% linked in the electronic edition to a data center may do so by tagging
%% their objects with \objectname{} or \object{}.  Each macro takes the
%% object name as its required argument. The optional, square-bracket 
%% argument should be used in cases where the data center identification
%% differs from what is to be printed in the paper.  The text appearing 
%% in curly braces is what will appear in print in the published paper. 
%% If the object name is recognized by the data centers, it will be linked
%% in the electronic edition to the object data available at the data centers  
%%
%% Note that for sources with brackets in their names, e.g. [WEG2004] 14h-090,
%% the brackets must be escaped with backslashes when used in the first
%% square-bracket argument, for instance, \object[\[WEG2004\] 14h-090]{90}).
%%  Otherwise, LaTeX will issue an error. 

%
%
\section{Introduction}

Novae are thermonuclear explosions in a layer of matter accumulated on the surface of a white dwarf
(WD) as a result of the accretion process from a companion star in the WD binary system. If the companion of the WD is a main sequence star then the nova is called a classical nova. In the rare case of a Red Giant (RG) companion, the nova is called symbiotic. It is expected that the recurrence time scale of nova explosions depends on the mass of the WD and the accretion rate. If the nova appears many times during a human life time 
in this same binary system, then it is called a recurrent nova. In this article we concentrate on the case of the symbiotic novae with a short recurrence periods (such as recently observed RS Oph).

The material expelled during recurrent nova explosion is slowed down due to the interaction with the surrounding medium forming a nebula, dubbed nova super-remnant (NSR). The structure of such NSR has been recently investigated by applying hydrodynamical simulations in the case of a specific recurrent nova M31N 2008-12a (in the Andromeda Galaxy) by 
Healy-Kalesh et al.~(2023). It is found that in subsequent eruptions a NSR is formed with the radius of tens of pc. In fact, such optical NSR has been recently detected with a projected size of at least 134 by 90 parsecs in the case of the most rapidly recurring nova M31N 2008-12a (Darnley et al. 2019).

Novae have been recently established as a new type of GeV and sub-TeV $\gamma$-ray sources \citep{2010Sci...329..817A, ack14, 2022Sci...376...77H,2022NatAs...6..689A},
indicating that particles (electrons, hadrons) are efficiently accelerated in their explosions. 
The GeV $\gamma$-ray emission is observed for several days to a few weeks after the nova explosion (e.g. Ackermann et al. 2014). However, for how long the process of acceleration of particles in nova shell is active is still an open issue. The  observed transient $\gamma$-ray emission has been 
proposed to be likely produced by hadrons in collisions with the matter expelled in a nova explosion 
(e.g. Tatischeff \& Hernanz~2007, Abdo et al. 2010, Sitarek \& Bednarek 2012, Martin \& Dubus 2013, Metzger et al. 2015, Ahnen et al. 2015, \citealp{2022NatAs...6..689A, 2022Sci...376...77H}, Cheung et al.~2022, 
Zheng et al.~2022).
However, relativistic electrons can also contribute (see e.g. Abdo et al. (2010), Sitarek \& Bednarek 2012, Martin \& Dubus 2013, Vurm \& Metzger~2018, Martin et al.~2018, Bednarek 2022), due to efficient energy losses on the IC process in the radiation from the nova photosphere or the companion RG star (in the case of symbiotic novae).    

Recently, Bednarek~(2022) considered the model in which the electrons can be accelerated in nova shells
for a much longer period of time than indicated by the $\gamma$-ray observations. In fact, provided that
acceleration of particles occurs within the nova shell, the acceleration process
could continue during the time scale corresponding to the recurrence period of the recurrent novae, at 
least in the part of the shell which propagates in the polar region of the nova binary system, where the shell
is not significantly decelerated. In terms of such a model,
the time dependent $\gamma$-ray emission produced in the Inverse Compton Scattering (ICS) of the radiation from the photosphere and the RG, as observed in the case of recurrent Nova RS Oph, has been calculated and compared with the sensitivities of the $\gamma$-ray telescopes. In \cite{bed22} it is concluded that with the future CTA this transient $\gamma$-ray emission from electrons in the shells should be detected during the period of the years.

In this follow-up work, we consider the fate of the relativistic electrons, which were accelerated in the nova shells but after some time escaped from them into the interstellar space. We suppose that such escaping electrons deposit in NSR around the nova. In the case of recurrent novae such NSRs should be able to accumulate large amount of relativistic electrons from multiple explosions. These electrons are expected to produce 
$\gamma$ rays by scattering mainly the Cosmic Microwave Background Radiation (CMBR) and thermal radiation from RG. 

As in Bednarek (2022), we consider the recurrent nova RS Oph which shows one of the shortest average recurrence period (on average $\sim$14.7 yrs, with the spread between 8.6 to 26.6  yrs), among the known Galactic recurrent novae see Tab. 21 in Schaefer (2010).
During its latest outburst in 2021 the source showed $\gamma$-ray emission from GeV energies (Cheung et al.~2022, Zheng et al.~2022) up to hundreds of GeV \citep{2022Sci...376...77H,2022NatAs...6..689A}, becoming the first nova detected in very-high-energy band. 
The RS Oph binary is composed of a White Dwarf (WD) and a RG with masses  1.2 -- 1.4$M_\odot$ and 0.68 -- 0.80 $M_\odot$, respectively \citep{2009A&A...497..815B}.
The same authors report an orbit with the period of 453.6 days and the inclination of 49$^\circ$ -- 59$^\circ$.
The surface temperature of the RG is estimated on 3600~K and its radius on $67^{+19}_{-16}$~R$_\odot$ (Dumm \& Schild~1998), where $R_\odot$ is the radius of the Sun .
The mass-loss rate of the RG is estimated on $5\times 10^{-7}M_\odot/\textrm{yr}$ \citep{2016MNRAS.457..822B}, however $3.7\times 10^{-8} - 10^{-6} M_\odot/\textrm{yr}$ values have been also suggested \citep{2008ASPC..401..115I, 2009ApJ...697..721S}. 
The wind concentrates around the equatorial plane of the binary system. The wind velocity is estimated on 40~km~s$^{-1}$ (Wallerstein~1958).
The upper limit on the mass expelled during nova explosion is constrained by the above mass loss rate of the RG and the recurrence period of the nova ($\sim 15$ yrs), on $5.6\times 10^{-7} - 1.5\times 10^{-5} M_\odot$.
This material forms very complicated structure around the nova. The observations of the nova RS Oph a few
hundred days after explosion in 2006 with the {\it Hubble Space Telescope} shows two-component flow, composed of a low-velocity high density equatorial region and a high-velocity low density polar region.
The polar region expended with the velocity of $(5600\pm 1100)$~km~s$^{-1}$ (Bode et al.~2006).
On the other hand, at about one day after the nova explosion, the expansion velocity was 
$\sim (4000-7500)$~km~s$^{-1}$ (Buil~2006). Therefore, the polar expansion seems not to decelerate significantly.
In the case of the outburst in 2021, the radio structure of the nova shell is very similar to that observed in 2006 (Munari et al. 2022). Also the velocity of the nova shell, averaged over the first 34 days, equal to $\sim$7550~km$^{-1}$~s$^{-1}$ is generally consistent with that observed in 2006.
The initial velocity of the shell is an important parameter in our modelling since it determines 
the ballistic time scale of the shell.

\citet{2008ApJ...688..559R} estimated the distance to RS Oph on $(2.45\pm0.37)$\,kpc 
(see also Bailer-Jones et al.~2021). This value is also very similar to the distance derived with the Gaia DR3-derived parallax: 
$2.68_{-0.15}^{+0.17}$~kpc (see also Schaefer~2022, Munari et al.~2022.
Multiple, partially contradicting distance measurements are available (see the discussion in \citealp{2022NatAs...6..689A}).

\section{Injection of electrons from the Nova}

In Bednarek (2022) we discuss the scenario in which electrons are assumed to be continuously accelerated 
in the nova shell long after nova explosion. Electrons, confined within the expanding shell, produce 
$\gamma$-rays by scattering RG and nova soft radiation.
here, we extend this model by assuming that electrons are finally realized from the shell into the nova surrounding. They diffuse in the NSR around recurrent nova producing persistent $\gamma$-ray emission. 

\subsection{Electrons within the shell}

It is assumed that electrons are accelerated within the shell region for a certain period of time, $t_{\rm inj}$, after the nova explosion. They are confined for some time within the expanding shell of the nova, losing energy on different radiation processes (i.e. synchrotron, IC, Bremsstrahlung). 
The shells move sub-relativisticly through the complex material surrounding the nova.
They are finally decelerated in the interaction with the interstellar matter.  
Relativistic electrons in the shell can be responsible for the extended $\gamma$-ray emission from novae on a time scale of years. The $\gamma$-ray emission produced by electrons within the shell on a time scale of years has been recently discussed in Bednarek~(2022).

\begin{figure}
\vskip 7.5truecm
\includegraphics{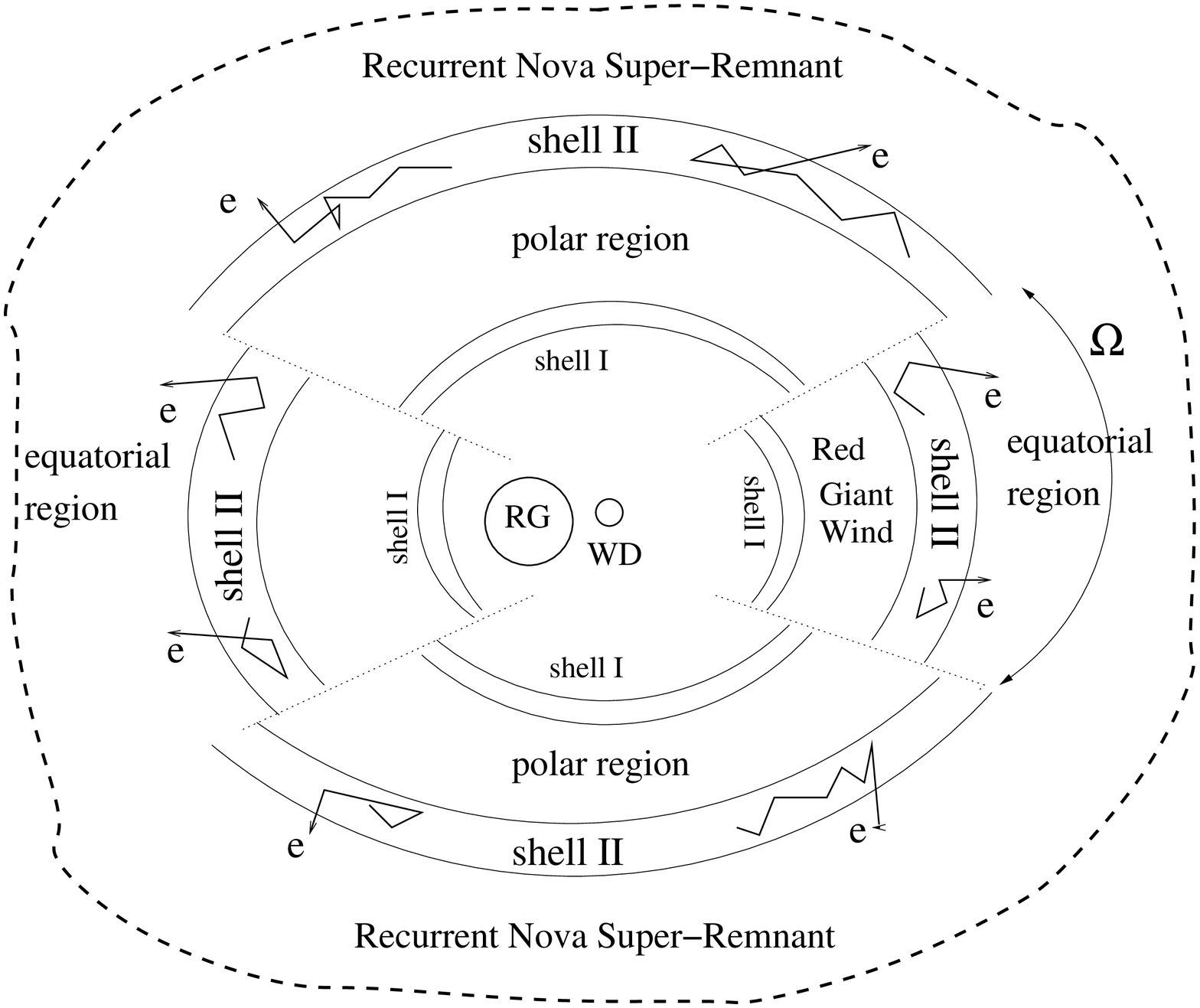}
\caption{Schematic picture of the NSR around a recurrent nova 
(not to scale). In the case of the Nova RS Oph, every recurrent explosion
injects a shell in which electrons are accelerated (2 shells are shown). The shells move with different velocities in the equatorial and polar regions of the binary system. A part of the sphere covered 
by the equatorial region is marked by the parameter $\Omega$.
In the equatorial region the shells are effectively decelerated by entraining the matter from the RG wind. 
Shells in both regions are decelerated on the sub-parsec distance scale by entraining the matter from the surrounding interstellar space. The electrons escape from the shells to the surrounding medium due to the (Bohm) diffusion process.
These electrons form a nebula around the recurrent binary system, subsequently producing $\gamma$-ray by ICS of the CMBR. 
The emission is expected from a region with a} size of the order of a few parsec at sub-TeV $\gamma$ rays.
\label{fig1}
\end{figure}

Up to our knowledge, there is no any specific model for the acceleration of particles in novae on a time scale of the nova recurrence period. All the presently considered models concentrate on the time scale
after the nova explosion corresponding to the observed time scale of the $\gamma$-ray emission already detected by the {\it Fermi}-LAT telescope, i.e. of the order of a month. Therefore, we introduce
a general model for the acceleration of the electrons, their energy losses, and their escape from the shell which is 
able to consider these processes on the time scale of the recurrence period of the nova. For the acceleration and propagation of electrons in the nova shell, we exploit the time-dependent model recently considered by Bednarek~(2022). Here we remind the main aspects of model. 
Two regions around the nova binary system with different proprieties are considered. Following other works \citep{2007A&A...464..119C,2008A&A...484L...9W, 2022ApJ...926..100M}, we assume that the wind from the RG is concentrated in the equatorial region of the binary system which extend is defined by its part of whole sphere (i.e. the parameter $\Omega$). A part of the nova shell, propagating in the equatorial region, is effectively decelerated due to the
entrainment of the matter from the RG wind. On the other hand, the (1 - $\Omega$) part of the nova shell, propagating in the polar region, expends freely up to the distances at which the matter entrained from the interstellar space decelerates the shell.
In fact such two component structure of the nova shell has been observed in the X-rays in the case of previous explosion of the Nova RS Oph see Montez et al.~(2022), in which case the fast shell takes about $\sim 0.2$ of the whole sphere and the slow shell $\sim 0.8$.
As a result of propagation in two different regions, their 
physical conditions (such as magnetic field strength, density of matter, radiation field, etc ...) differ significantly. These conditions determine the acceleration process of the electrons, their energy losses in the shell and also the escape conditions from the shell into the NSR. 

The electrons are assumed to be injected into the nova shell with a power-law spectrum and an exponential cut-off during initial time interval starting from the nova explosion.
We develop a simple model which allows to determine the basic physical parameters of the expending shell
in the long period after explosion such as, its expansion velocity and density of matter, the magnetic field strength, 
and different radiation fields (see for details Sect.~3 in Bednarek 2022). 
The magnetic field is assumed to be at some level of equipartition with the kinetic energy of the shell.
We take into account the radiation field from the RG and also from the nova photosphere at the early time after explosion.
Then, we define the acceleration process of electrons, as a function of time, and their energy losses.
The cut-off energies in the electron spectrum are obtained by balancing the acceleration time scale with their time scale for energy losses or the dynamical time scale of the shell.
The total power of the electrons is normalized to the initial kinetic energy of the shell. Since the electrons lose energy on different radiation processes, we numerically calculate the evolution of the equilibrium spectrum of the electrons within the shell at an arbitrary moment after explosion (for details see Sect.~4 in Bednarek~2022).
Due to the time-dependent conditions within the shell, we follow the fate of electrons with different energies applying the time step method to a single shell.
With $\Delta t$ being the time step, the location of the shell at the time 
$t_{\rm n}$ is calculated from, $R_{\rm n} = R_{\rm n-1} + v_{\rm sh}(t_{\rm n-1})\cdot \Delta t$.   
The thickness of the shell is related to its radius according to $\Delta R = \beta R$, with $\beta<1$. In such a model, we define the condition for the escape of the electrons from the shell into the interstellar region. We assume that the diffusion process of the electrons is well described by the Bohm's prescription. However, since the physical conditions in the propagating shell vary in time, we calculate the characteristic diffusion distance of electrons in the shell from
\begin{eqnarray}
R^{\rm dif}_{\rm n} = R^{\rm dif}_{\rm (n-1)} + \sqrt{D_{\rm B}/(2t_{\rm n})}\cdot \Delta t,
\label{eq10}
\end{eqnarray}
\noindent
where the Bohm diffusion coefficient is $D_{\rm B} = cR_{\rm L}/3$, and 
$R_{\rm L} = E/(eB)\approx 3\times 10^{16}E_{10}/B_{\mu G}$~cm  is the Larmor radius, $E_{\rm e} = 10E_{10}$~TeV is the electron energy, and $B =10^{-6}B_{\rm \mu G}$~G is the magnetic field strength at the
time  $t_{\rm (n-1)}$.
The electrons can escape from the shell into the interstellar medium when the diffusion distance becomes comparable to the thickness of the shell, 
i.e. $R_{\rm dif} = \beta R$.

\subsection{Electrons released from the shell into the NSR}

In the present work we consider the fate of electrons which were able to escape from the nova shells.
We argue that these relativistic electrons accumulate in nova surrounding. 
The electrons are confined around the nova for a certain time, diffusing slowly in the outward direction into the NSR.
We investigate the $\gamma$-ray emission produced by the electrons in the IC scattering 
of the radiation from the RG (a companion of the WD),
and the CMBR. 
We include the energy losses of the electrons within the shell on the ICS of the thermal radiation from the nova photosphere during its initial propagation for the period of about 1 month.
However, averaged over the whole recurrence period of RS Oph, this photosphere radiation field is negligible when compared to the RG thermal radiation, hence it is not taken into account when computing the persistent $\gamma$-ray production in the nova super-remnant.
Our interest in the existence of relativistic electrons in the NSR we focus on the case of the recently detected in the sub-TeV range
 Nova RS Oph \citep{2022NatAs...6..689A, 2022Sci...376...77H}. 
In the case of RS Oph six -- nine\footnote{Three out of those nine outbursts were only indirectly observed.}  explosions have been documented with the recurrence period of the order of 15 years. Those observations indicate that electrons can be accelerated to TeV energies at least in the early period after the nova explosion.
The schematic picture of the scenario considered here is shown in Fig.~1.

Following the method described in the previous subsection for the evaluation of processes within the nova shell, we calculate the spectra of the electrons escaping from the shell into the interstellar medium for the parameters of the nova RS Oph and reasonable parameters of the acceleration model of the electrons.
We consider the electron injection spectra with the spectral index -2 (Bell~1978).
In Fig.~2 we show how the maximum energies of the electrons evolve in time for a few different values of the magnetization of the shell. The magnetization parameter of the shell, $\alpha$, which is obtained assuming some level of the equipartition between the kinetic energy density of the shell and energy density of the magnetic field in the shell, i.e.  $\alpha n_{\rm sh} m_{\rm p} v_{\rm sh}^2/2 = B^2/(8\pi)$, where  
$n_{\rm sh}$ is the density of the matter in the shell, $m_{\rm p}$ is the proton mass, $v_{\rm sh}$ is the shell velocity, and $B$ is the magnetic field strength within the shell (see for details Bednarek 2022). 
Both cases, with and without deceleration in the RG wind, are considered. 
The maximum energies of the electrons are limited either by the energy losses (at the early time) or the dynamical time scale of moving shell (at the latter time).
They become larger at latter time due to decreasing magnetic field within the shell. These maximum energies  are also significantly lower in the case of drastically decelerated shell since the acceleration parameter is assumed to depend on the shell velocity (see Eq.~3 in Bednarek 2022). In this section we consider a relatively slow acceleration process of the electrons in the shell with the energy gain proportional to 
$\xi\sim (v_{\rm sh}/c)^2$ (where $v_{\rm sh}$ and $c$ are speed of the shell and light respectively), which is characteristic for the second order Fermi acceleration process. 
When calculating the $\gamma$-ray emission from nova super-remnants, we will discuss also more efficient acceleration process. 

\begin{figure*}
\vskip 5.5truecm
\includegraphics{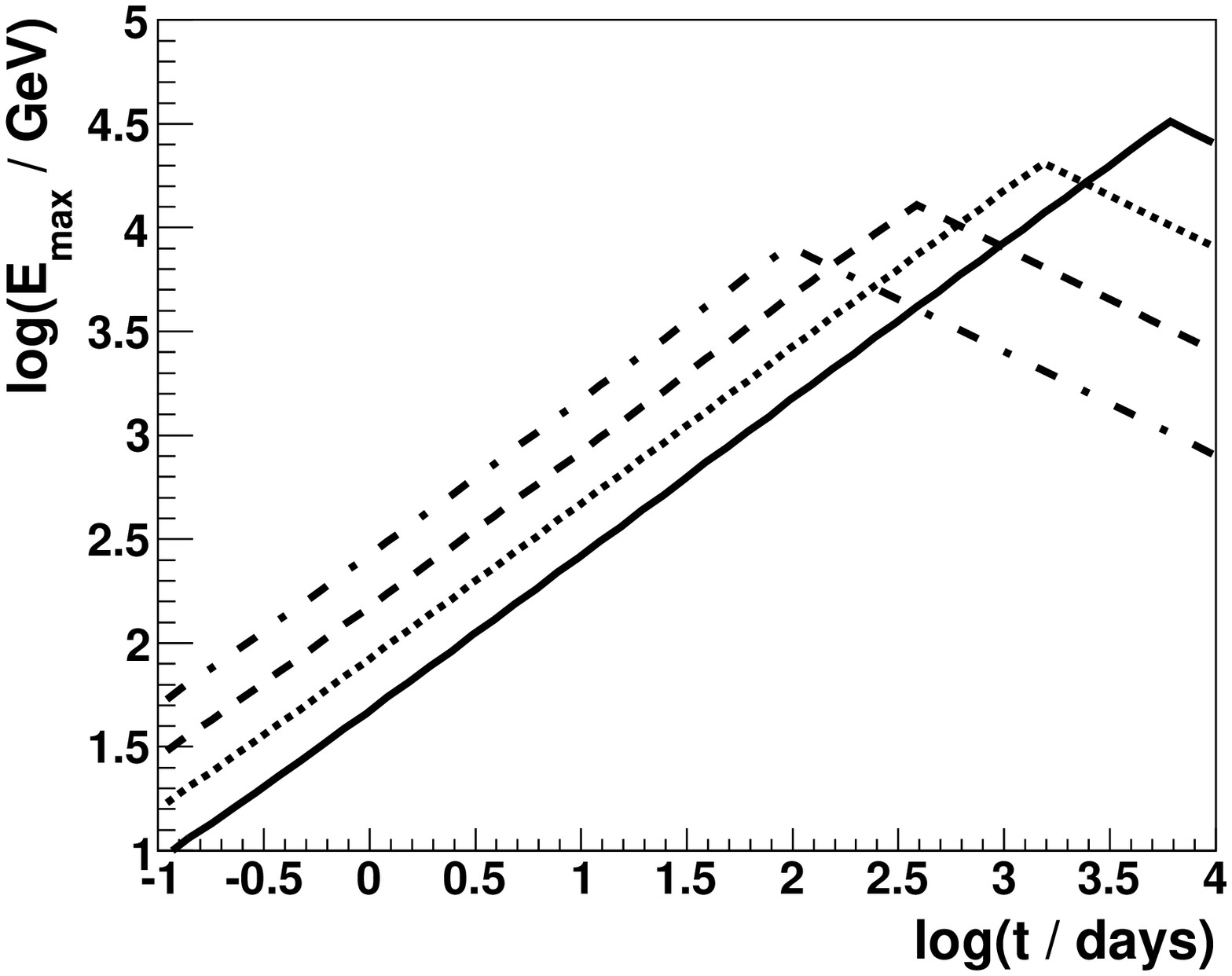}
\includegraphics{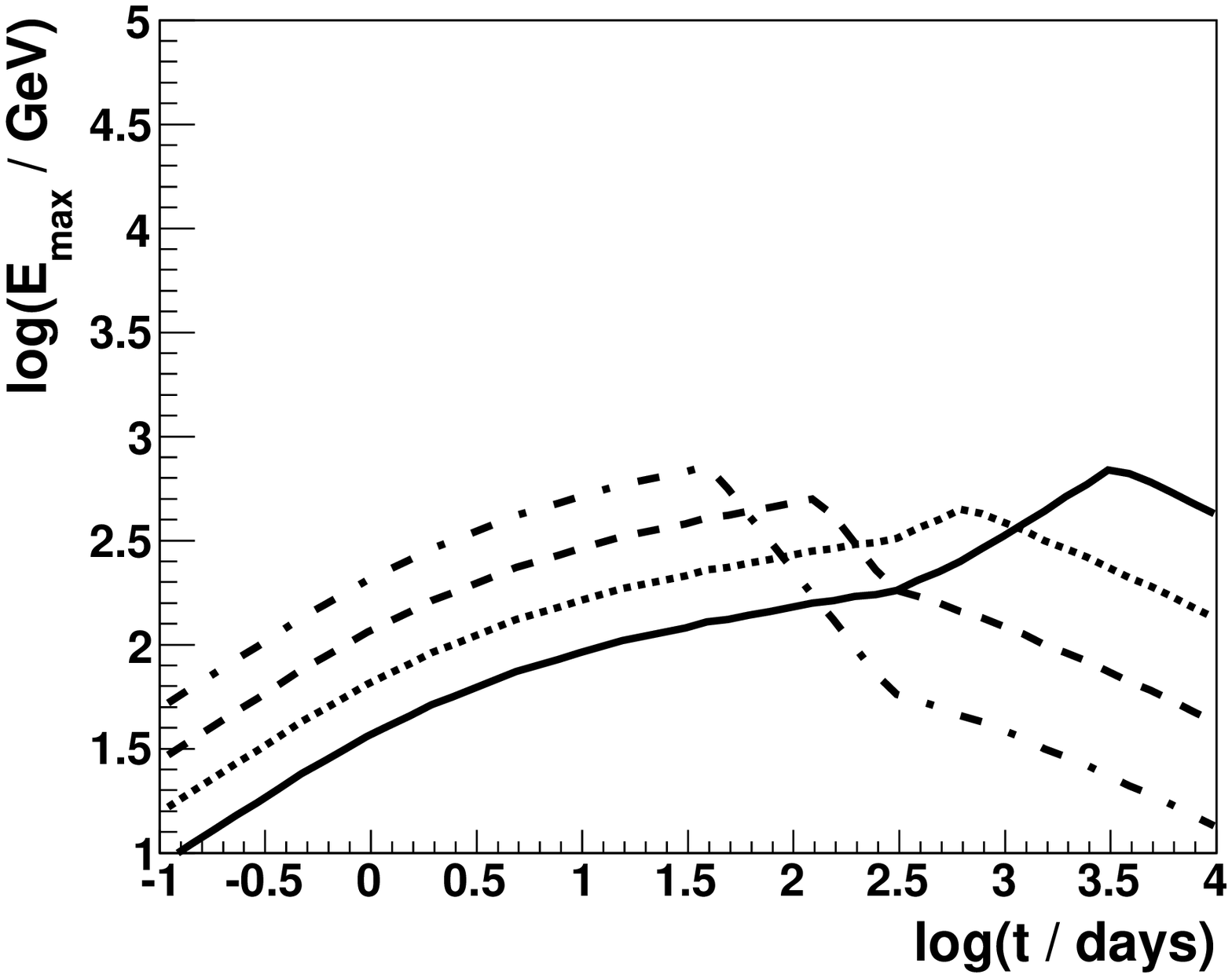}
\includegraphics{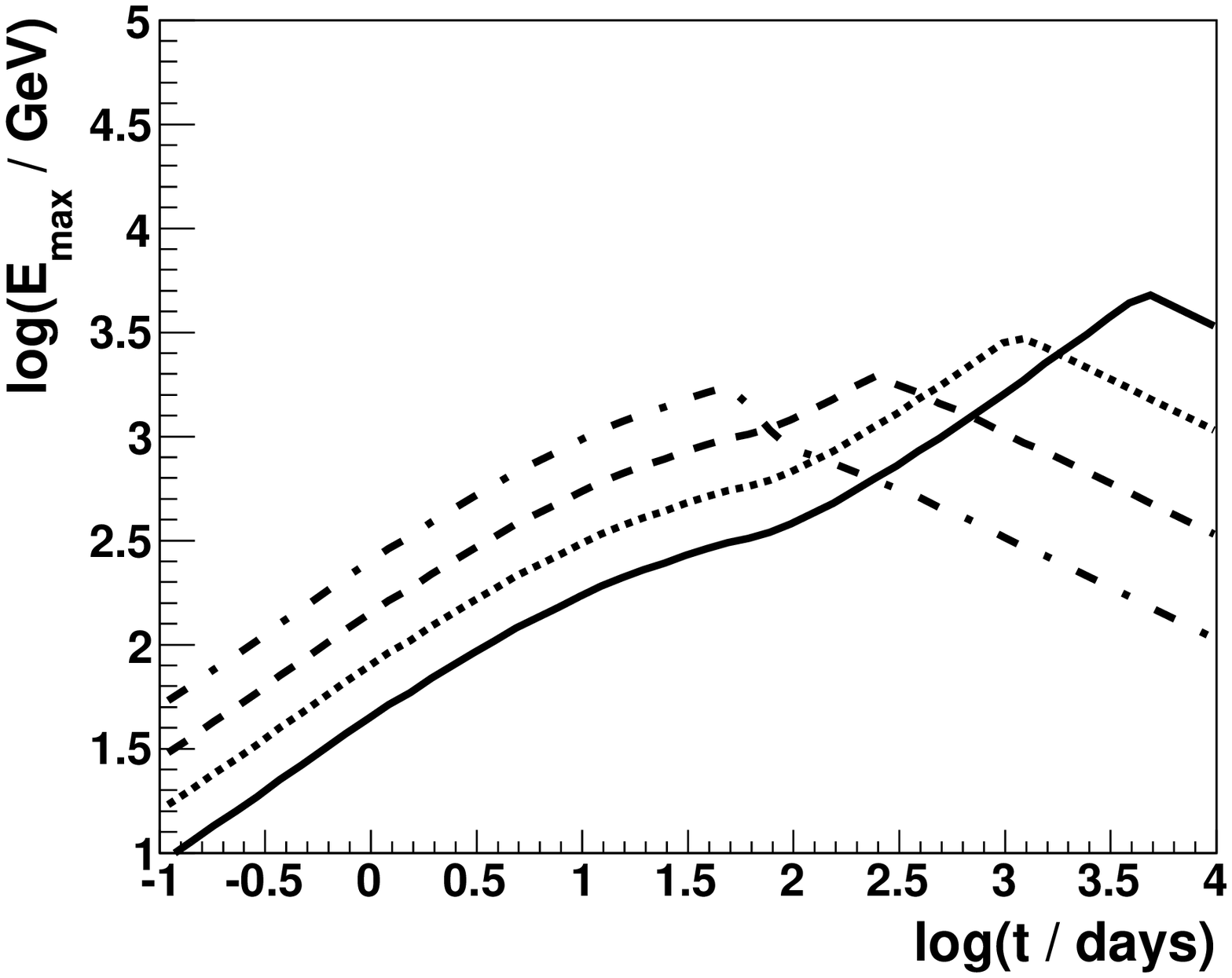}
\caption{The maximum energies of the electrons in the shell as a function of propagation time of the shell. The electrons are accelerated within the nova shell in the case of its propagation in
the polar region of the binary system, the so-called ``no deceleration'' case, (see panel on the left) and in the case of propagation in the equatorial 
wind of the RG (deceleration case) which is confined within the part of the whole sphere defined by 
$\Omega = 0.1$ (in the centre) and 0.5 (on the right). 
The magnetization of the nova shell is defined by $\alpha = 10^{-4}$ 
(dot-dashed curve), $10^{-3}$ (dashed), 0.01 (dotted), and 0.1 (solid).
The initial mass of the nova shell is 
$M_{\rm sh}^{0} = 10^{-6}~M_\odot$ and its initial velocity is $v_{0}^{\rm sh} = 6000$~km~s$^{-1}$.
The mass loss rate of the RG wind is $M_{\rm RG} = 10^{-7}$~M$_\odot$~yr$^{-1}$ and its velocity 
$v_{\rm RG} = 40$~km~s$^{-1}$.
}
\label{fig2}
\end{figure*}

We apply a simple model for the initial acceleration of the electrons since our main goal is to consider a radiation model for the nova super-remnant (formed by electrons escaping from the shell) in the late stages. A few more detailed models for the high energy processes in the initial stage after a nova explosion have been considered already.   
For example, Tatischeff \& Hernanz (2007) consider non-linear acceleration of particles at the initial blast wave and show that protons can reach TeV energies already at days time scale after explosion. 
Also Martin \& Dubus~(2013) develop a model for the shock acceleration and radiation of electrons and protons 
in the shocked plasma during initial month time scale after the nova explosion. Martin et al.~(2018) consider detailed model for the acceleration of particles in collisions of the wind from the WD with the initial shell of the nova during first month after explosion.  
They show that $\gamma$-ray emission comes mainly from hadronic interactions with the dense matter downstream of the shock. Similar general model, but limited to the equatorial region of the nova binary system, has been also considered by Metzger et al.~(2015). Detailed calculations of the $\gamma$-ray emission in terms of the hadronic and leptonic scenario, during initial stage of nova, are presented in Vurm \& Metzger~(2018). 
All these models have been developed in order to explain the $\gamma$-ray emission from the initial stage of the nova explosion as observed by the {\it Fermi}-LAT during the time scale of the order of a month.
Here we consider a model for possible $\gamma$-ray emission from novae at long time after explosion during the whole activity period of the recurrent nova.
We argue that electrons, escaping from the shell can accumulate for a long time in the vicinity of the recurrent nova. They produce $\gamma$-ray nebula similar to those observed around rotation powered pulsars.

We investigate the dependence of the spectrum of the escaping electrons on the basic parameters of the model, assuming the likely parameters of the nova RS Oph. In Fig.~3 we show the spectra of the electrons escaping from the polar part of the shell, assuming that they are accelerated only during short time (30 days) after 
the explosion of the nova or during the whole recurrence period of RS Oph (the average period of 15 yrs is assumed). The shorter time scale (of the order of a month) is consistent with the observed by the 
{\it Fermi}-LAT telescope time scale for the initial GeV gamma-ray emission from Nova RS Oph 
(Cheung et al. 2022). We suppose that the acceleration process of electrons lasts at least for
the period of the initial $\gamma$-ray production phase as observed by the Fermi-LAT.
Note also, that such order of the time scale for acceleration of particles is expected
in the model by Martin et al. (2018) in which the acceleration process is active during the time of collision of the fast wind from the white dwarf with a slower moving initial shell of the nova. 
The longer time scale (of the order of the recurrence period of the Nova RS Oph ~15 yrs) is
motivated by the fact that at this time scale the nova shell is expected to significantly 
decelerate due to the entrainment of the interstellar matter. In order to estimate the deceleration distance of the shell we introduce a model for the velocity of the shell as a function of distance
from the nova. We assume a conservation of the momentum 
of the shell interacting with the surrounding medium, 
$M^{\rm sh}_{\rm 0}v^{\rm sh}_{\rm 0}  = v_{\rm sh}(R)[M_{\rm sh} + M_{\rm cos}(R)]$,
where $M^{\rm sh}_{\rm 0}$ and $v^{\rm sh}_{\rm 0}$ are the initial mass of the nova shell and its initial velocity, $M_{\rm cos} = (4/3)\pi R^3 n_{\rm cos}m_{\rm p}$ is the mass entrained from the interstellar medium, $R$ is the radius of the shell, 
$v_{\rm sh}(R)$ is the velocity of the shell at the distance $R$, and $n_{\rm cos}$ is the density of the interstellar medium.
For reasonable densities of the interstellar medium (in the range $n_{\rm cos} \approx 0.1 - 10$~cm$^{-3}$), the shell starts to significantly decelerate at the distance of the order of $\sim 10^{17}$ cm (see the results of calculations presented in 
Fig. 6 in Bednarek 2022). This distance scale is passed by the shell during the time similar to the recurrence period of the Nova RS Oph. Our simple model for the velocity profile generally agrees 
with the results of observations for the outburst of nova RS Oph during the early expansion phase reported 
in Tatischeff \& Hernanz~(2007, see Fig.~1). In our model (described by the solid curve in Fig.~6a, Bednarek 2022) the velocity of the forward shock, moving in the polar region, drops by a factor of the order ~2 during the first $\sim$30 days as also observed in the case of the outburst in RS  Oph during 2006 
(Tatischeff \& Hernanz~2007).  

As we have shown above, the efficiency of the acceleration process of electrons, 
$\xi$, depends in our model on the velocity of the shell. Therefore, it has to drop significantly at the time scale of the order of the recurrence period of the Nova RS Oph.
Interestingly, there are observational evidence for shocks appearing at late stages after nova explosion.
\citet{2022MNRAS.515.3028B} reported enhancement of radio emission starting 1.5 months after the eruption, and lasting at least for a month more. It is interpreted as synchrotron emission due to internal shocks within the ejecta. Moreover, at least 25\% of novae observed by Chomiuk et al. (2021) show  non-thermal synchrotron emission. Two of them, V5589 Sgr and V392 Per, appear to be on the bridge between classical and symbiotic novae. Another example of the late non-thermal synchrotron emission (months time scale) is the
symbiotic recurrent nova V3890 Sgr (Nyamai et al. 2023).  

The cases of slow and fast acceleration of the electrons within the shell are considered. They depend on the velocity of the shell. In the case of fast acceleration the acceleration coefficient is 
assumed to be $\xi\sim (v_{\rm sh}/c)$. Specific spectra of the escaping electrons depend on the 
magnetization parameter of the shell which determines the diffusion process of the electrons in the shell and one of the main energy loss processes (i.e. synchrotron emission). 
We conclude that long acceleration process of the electrons results in their spectra extending to higher energies. In the case of fast acceleration scenario, the electrons are able to escape into NSR with multi-TeV energies. In the case of slow acceleration, typical energies of the escaping electrons are one-two orders of magnitude lower. The spectra of the electrons show interesting dependence on the magnetization parameter of the shell. For extended acceleration (during 15 yrs),
the spectra of the electrons extend to larger energies for strong magnetization. This is due to the fact that the maximum energies of the electrons are determined in this case by the dynamical time scale of the shell.  
For acceleration limited to the early stage of the shell propagation, the dependence of the electron spectra on the magnetization is just the opposite. This is due to the fact that, soon after the nova explosion, the maximum energies of the accelerated electrons are determined by the synchrotron energy losses.
This complicated behaviour is also in line with the dependency of the maximum electron energies inside the shell shown in Fig.~2.

\begin{figure*}
\vskip 5.5truecm
\includegraphics{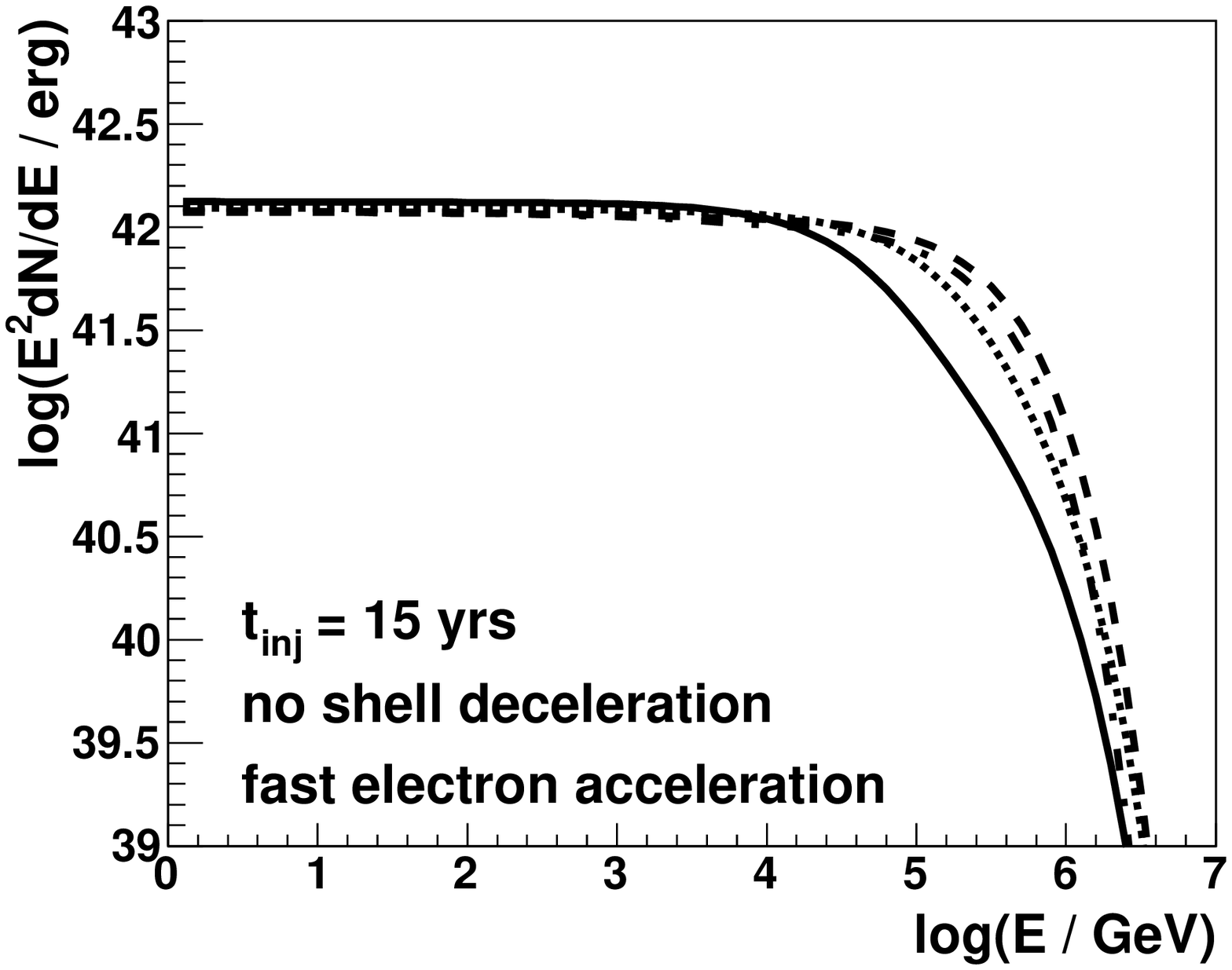}
\includegraphics{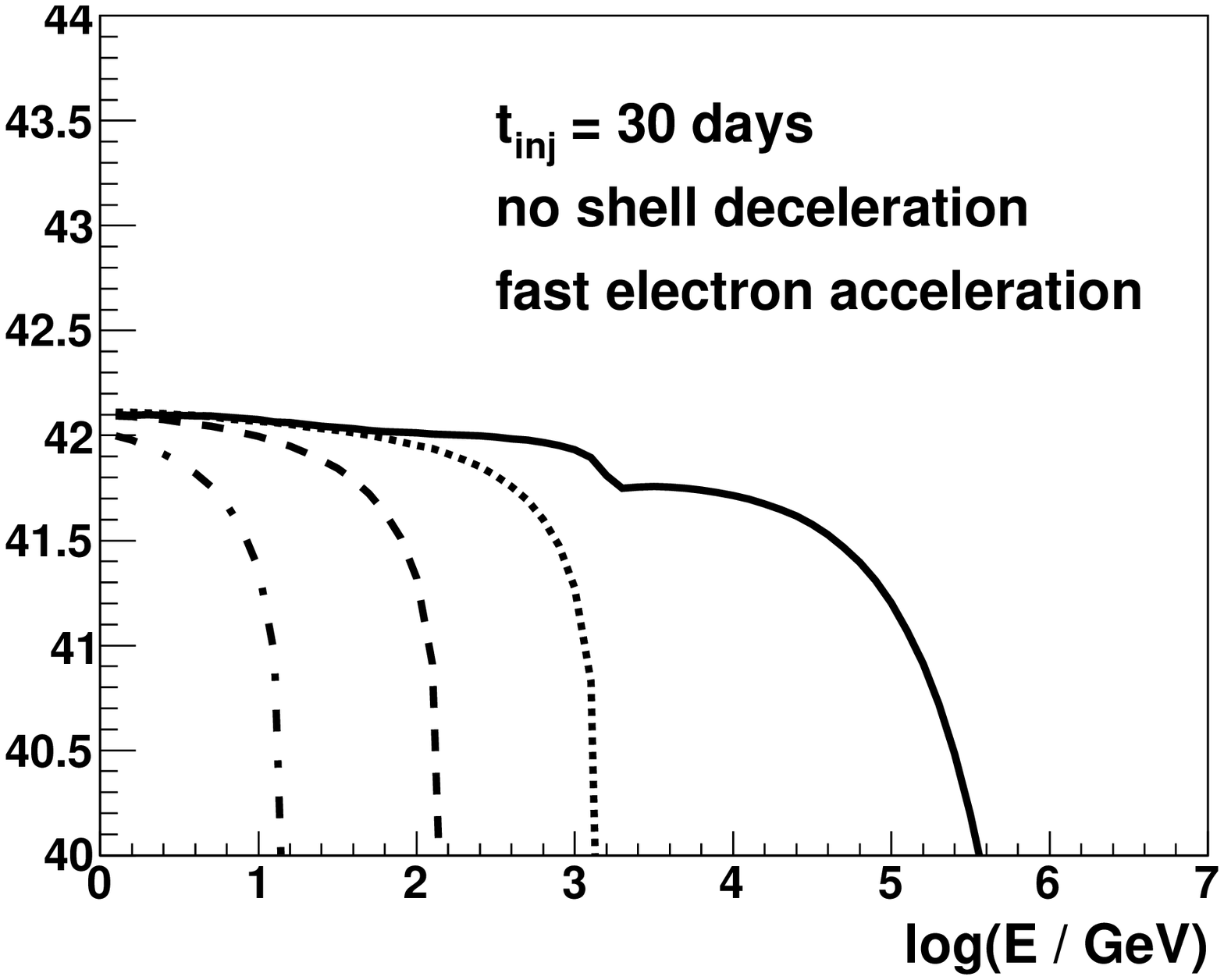}
\includegraphics{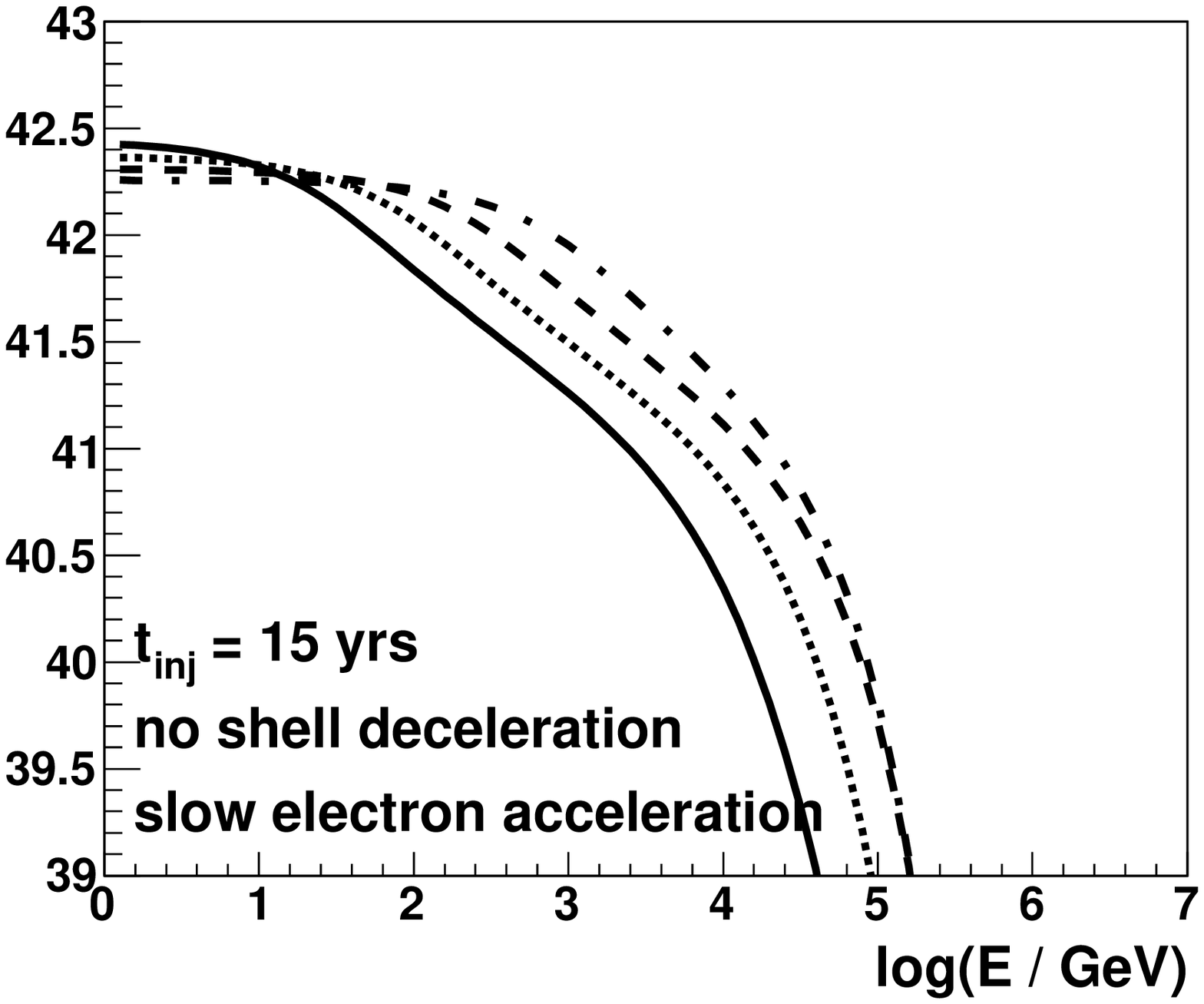}
\includegraphics{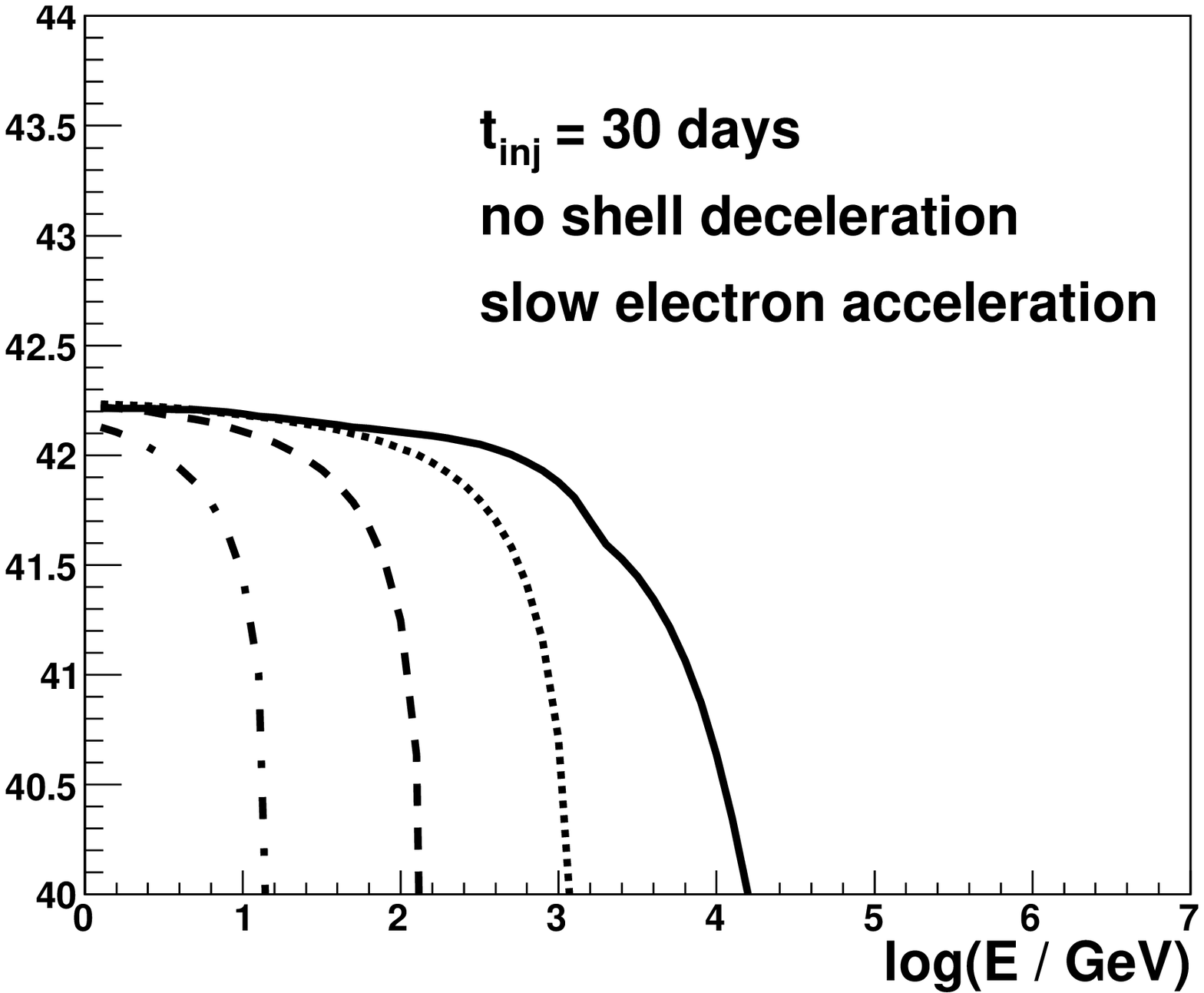}
\caption{The differential spectral energy distribution of the electrons escaping from the nova shell into the NSR assuming that their acceleration process operates only during the short period 
after the nova explosion (assumed to be 30 days) or during the whole recurrence period of the nova 
(i.e. 15 yrs). The electrons are accelerated within the shell to a power-law spectrum with a spectral index $-2$ and the exponential cut-off at $E_{\rm max}$. The cut-off energy of  $E_{\rm max}$ is defined by two acceleration prescriptions, slow and fast (see text for details). 
The magnetization of the nova shell is defined by $\alpha = 10^{-5}$ 
(solid curve), $10^{-4}$ (dotted), $10^{-3}$ (dashed), and $10^{-2}$ (dot-dashed).
The shell propagates in the polar region of the nova binary system (no deceleration case).
The initial mass of the nova shell is 
$M_{\rm sh}^{0} = 10^{-6}~M_\odot$ and its initial velocity is $v_{0}^{\rm sh} = 6000$~km~s$^{-1}$.
}
\label{fig3}
\end{figure*}
\begin{figure*}
\vskip 5.5truecm
\includegraphics{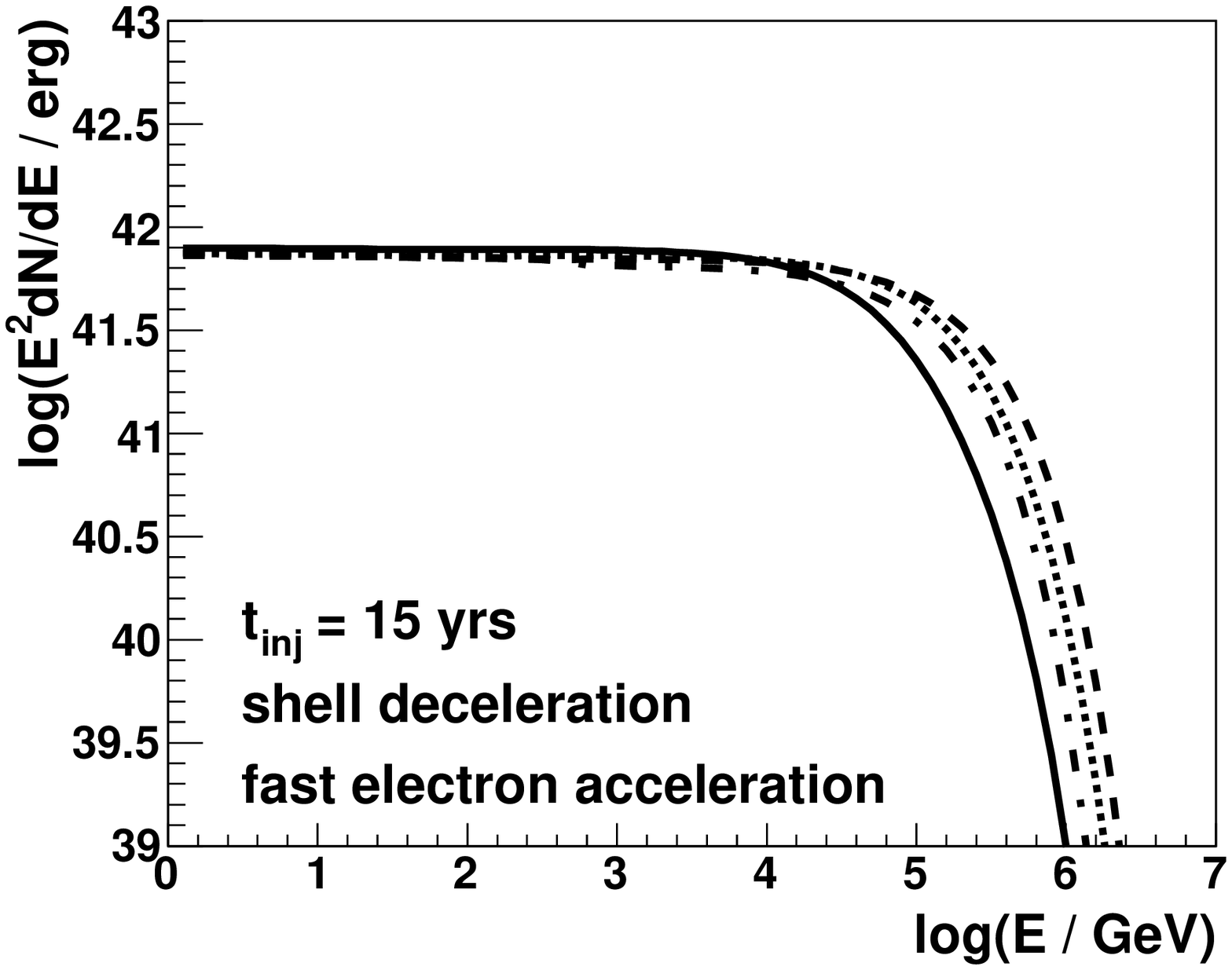}
\includegraphics{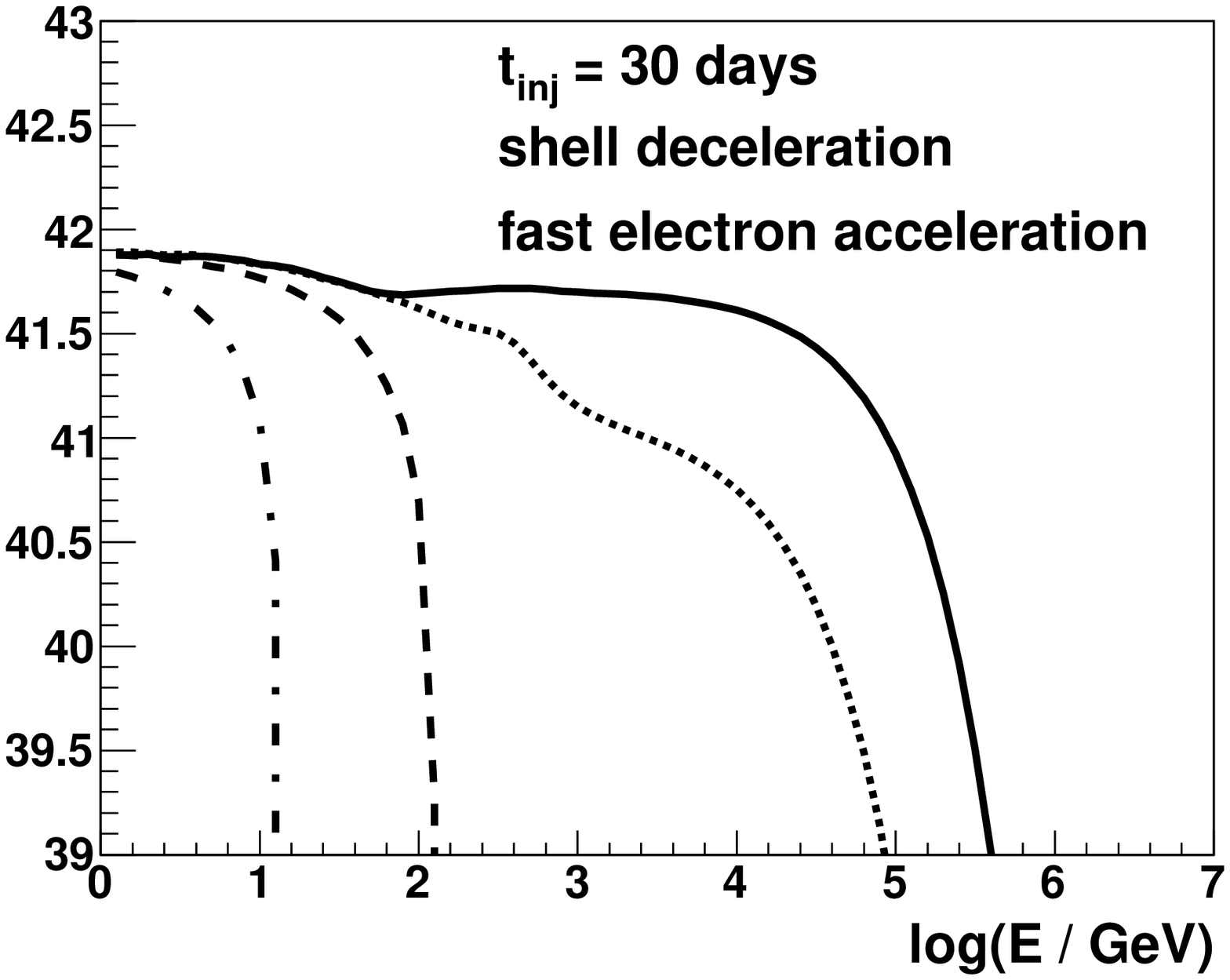}
\includegraphics{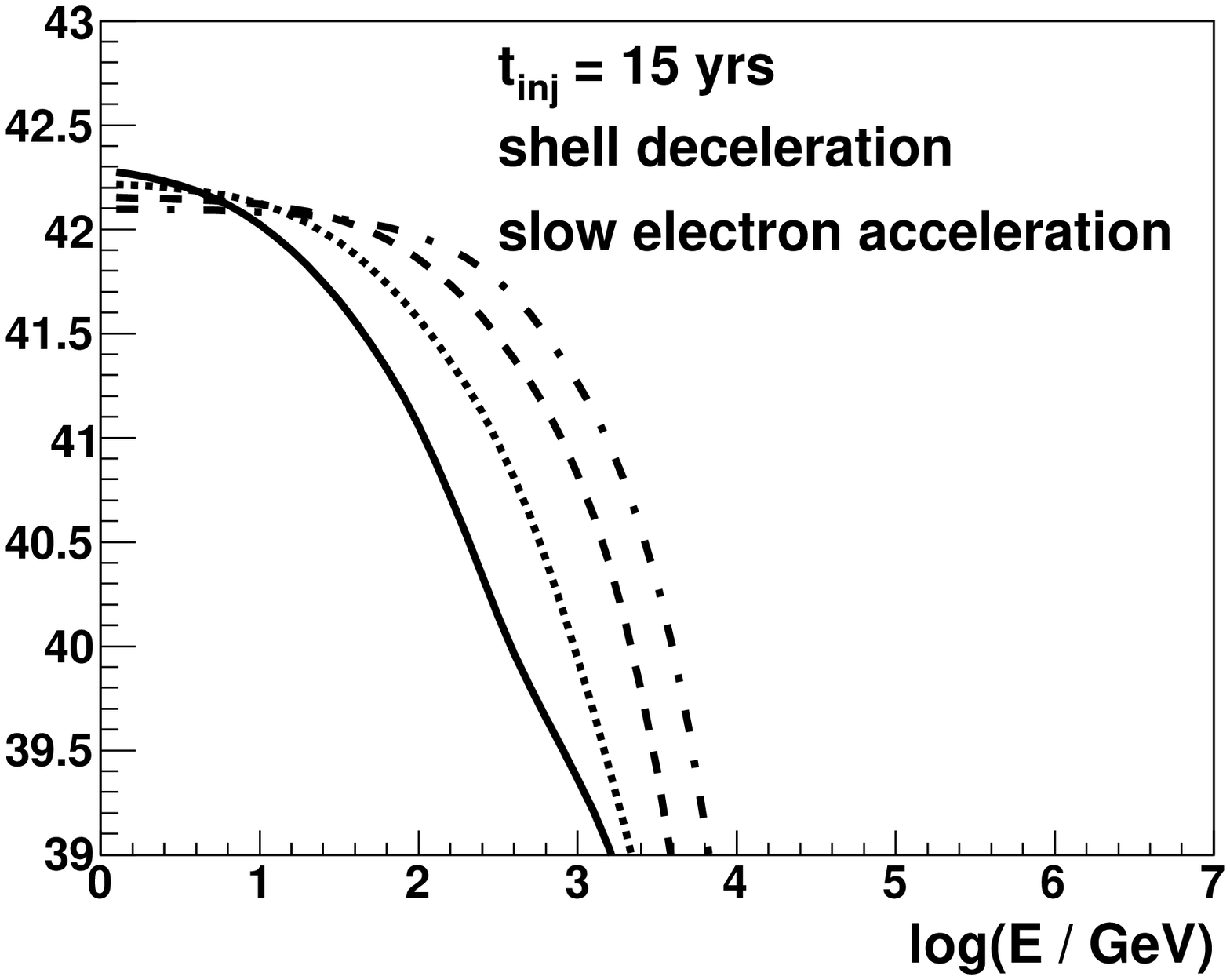}
\includegraphics{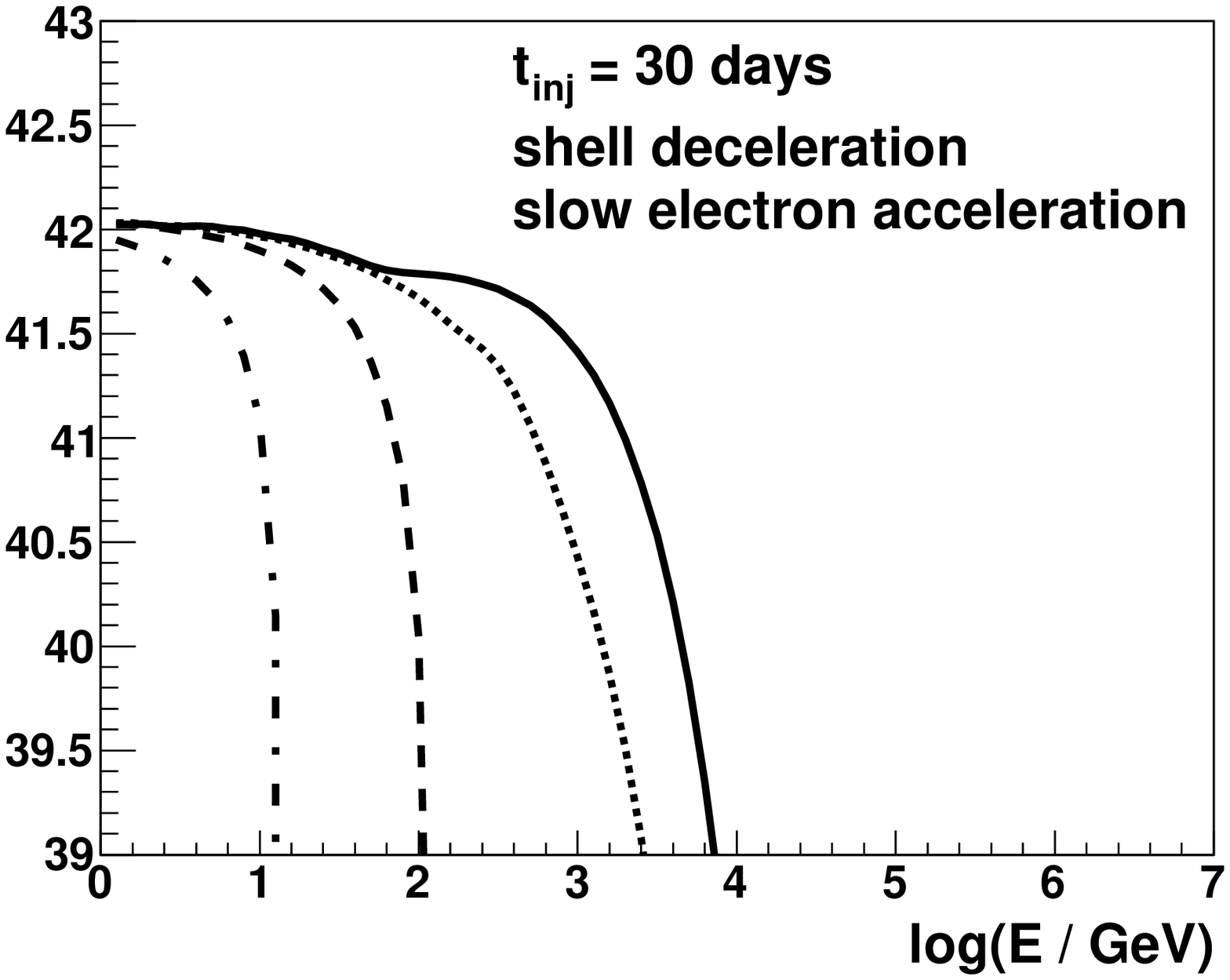}
\caption{As in Fig.~3 but for the part of the shell propagating in the wind of the RG expelled in the equatorial region of the nova binary system. Then, the shell decelerates according to the prescription defined in Sect.~6 in Bednarek~(2022). A part of the sphere in which the shell propagates is fixed on 
$\Omega = 0.3$.}
\label{fig4}
\end{figure*}

In Fig.~4, we investigate the spectra of the escaping electrons in the part of the shell which propagates in the equatorial region of the nova binary system.
Since this part of the shell decelerates fast, due to the entrainment of the dense RG wind material, the energies of the accelerated electrons are lower by one-two orders of magnitudes. Moreover, the electrons in the shell have much more time to lose energy on radiation processes already during the confinement within the shell. Therefore, their energies are usually limited to the sub-TeV range. The electrons escaping with such energies into the nova super-remnant mainly contribute to the $\gamma$-ray emission in the GeV range.

Note that recurrent novae of the RS Oph type, are expected to produce many explosions in a relatively short 
period. In fact, the electrons escaping from the last shell should pass through the content of the
previous shells. This process is difficult to consider due to the unknown structure and magnetization of the previous shells
at some time after slow down. We expect that  such ``old'' shells affects the magnetic field strength
within the inner part of the NSR. We consider this effect by studying much stronger magnetic field within the NSR (than expected from the values of the interstellar magnetic field strength) in the further sub-section. 

It is assumed that nova shells decelerate completely and merge with the interstellar medium at the distance $R_{\rm inj} = 2\times 10^{17}$~cm, which is a typical distance for the deceleration of the shells in the medium with density of the interstellar space, equal to 1 particle~cm$^{3}$ (see Fig.~6 in Bednarek~2022).
Thus, all the electrons that are still present in the shell at this time feed the NSR.
For simplicity we assume that the electrons that were able to escape earlier as well diffuse in the interstellar medium to $R_{\rm inj} = 2\times 10^{17}$~cm, hence the injection of all the electrons into the NSR occurs at the same distance.
In most of the cases this is a reasonable assumption.
However, the electrons that have energies $\lesssim 10$~GeV and escaped closer than $3\times10^{16}$~cm would not be able to diffuse in the magnetic field of 3~$\mu$G up to $R_{\rm inj}$ due to IC energy losses on RG radiation field.
Nevertheless such low energy electrons are also not likely to diffuse away out of the shell having still strong magnetic field, and would more likely by trailed by it up to dispersion of the shell at $R_{\rm inj}$. 
Due to Klein-Nishina effect higher energy electrons suffer smaller energy losses during the diffusion and e.g. 1~TeV electrons can keep most of their energy even if they escape from the shell at the distance of $3\times10^{15}$~cm. 

\section{Gamma rays from NSR around a recurrent nova}

As we have shown above, the acceleration process of the electrons depends on the conditions within the nova shell. In the case of recurrent novae, many shells are expected to provide relativistic electrons to the medium surrounding the nova super-remnant.
The electrons finally escape from the nova shell into the surrounding medium. They produce $\gamma$ rays in the IC process by scattering the CMBR and optical radiation from the RG. 
We distinguished two main regions in which a nova shell propagates, the equatorial wind region and the polar region. The physical conditions for the electrons in these two regions differ significantly. Electrons are injected into the NSR from these two regions producing $\gamma$-ray emission with features which can differ significantly. 
In this section we investigate the $\gamma$-ray spectra (and their detectability) produced by the electrons from these two regions as a function of the free parameters of the model.
As an example, we consider the physical parameters of the nova RS Oph and its binary system (see Introduction). We assume that this recurrent nova explodes with the average recurrence period of 15 yrs for the last  $t_{\rm max} = 10^4$ yrs or $10^5$~yrs. This time scale for the activity period of the recurrent nova determines the total energy transferred to the relativistic electrons in the nova super-remnant.

Relativistic electrons, which were able to escape from the nova shells, diffuse into the surrounding medium forming NSR. It resembles well known nebulae around rotation-powered pulsars.
During the diffusion process electrons lose energy on the synchrotron and the Inverse Compton processes.
They produce $\gamma$ rays by scattering the CMBR and RG radiation.
In order to follow the fate of the electrons in the nova super-remnant we 
modified the Monte Carlo code which has been originally 
developed for the diffusion of electrons within globular clusters (see Bednarek \& Sitarek 2007, and 
Bednarek et al. 2016). The code follows the diffusion process of electrons through the globular cluster and their energy losses on the synchrotron process and the inverse Compton process in the optical radiation from the stars and the CMBR. In the code, we replaced the radiation field from the stars in the globular cluster by
the radiation field of the RG.
In this way we are able to calculate the $\gamma$ ray (and also synchrotron) spectra produced by the relativistic electrons during the activity time of the recurrent nova.

\subsection{Contribution from the polar parts of the shells}

In Fig.~5, we show the $\gamma$-ray spectra produced in the nova super-remnant by the electrons accelerated in terms of the two acceleration scenarios. They are defined by different prescriptions for the acceleration coefficient
which defines the energy gain rate of the electrons, $\xi= v_{\rm sh}/c$ (fast acceleration, see figures (a),(b), (e) and (f)) and $\xi = 0.1(v_{\rm sh}/c)^2$ (slow acceleration, see figures (c), (d), (g) and (h), see also Sect.~3, Bednarek 2022).
It is assumed that the electrons are injected continuously during the recurrence period of the nova.
The spectra are shown for different magnetization parameter of the nova shells, $\alpha$. 
Two activity periods of the recurrent nova are considered, $T_{\rm active} = 10^4$~yrs and $10^5$~yrs 
(see Fig. 5). 
In the case of fast acceleration, the $\gamma$-ray spectra clearly extend through the multi-TeV energy range,
with slow dependency on the value of $\alpha$. 
They should be easily detected by the CTA. The extensive observations with the present Cherenkov telescopes such as H.E.S.S., MAGIC, or VERITAS can also significantly constrain the allowed range of considered parameter space of the model.
As RS Oph has a rather low Galactic latitude 
(Galactic coordinates $l = 19.8^\circ$, $b \sim 10.37^\circ$), where the sensitivity of GeV instruments strongly varies, we compare the expected emission with two bounding, 
publicly-available\footnote{\protect\url{https://www.slac.stanford.edu/exp/glast/groups/canda/ lat\_Performance.htm}} sensitivities: for latitudes  $b = 0^\circ$ and $30^\circ$.
It appears that the sensitivity line for the nova RS Oph location should stay somewhere in the middle between these two (marked) lines.
Therefore, we conclude that
in the case of the older nova super-remnants (with the activity time equal to $T_{\rm active} = 10^5$~yrs, see Figs. 5b,d), the $\gamma$-ray emission
is expected to be also constrained by the \textit{Fermi}-LAT telescope at GeV $\gamma$-ray energy range. In the case of the slow acceleration model for electrons, the $\gamma$-ray emission is limited to sub-TeV energies (see Fig.~5cd).
In this case, $\gamma$ rays can hardly be detected by the CTA. But, the allowed parameter space of the model is expected to be likely constrained by the observations with the \textit{Fermi}-LAT.

\begin{figure*}
\vskip 8.truecm
\includegraphics{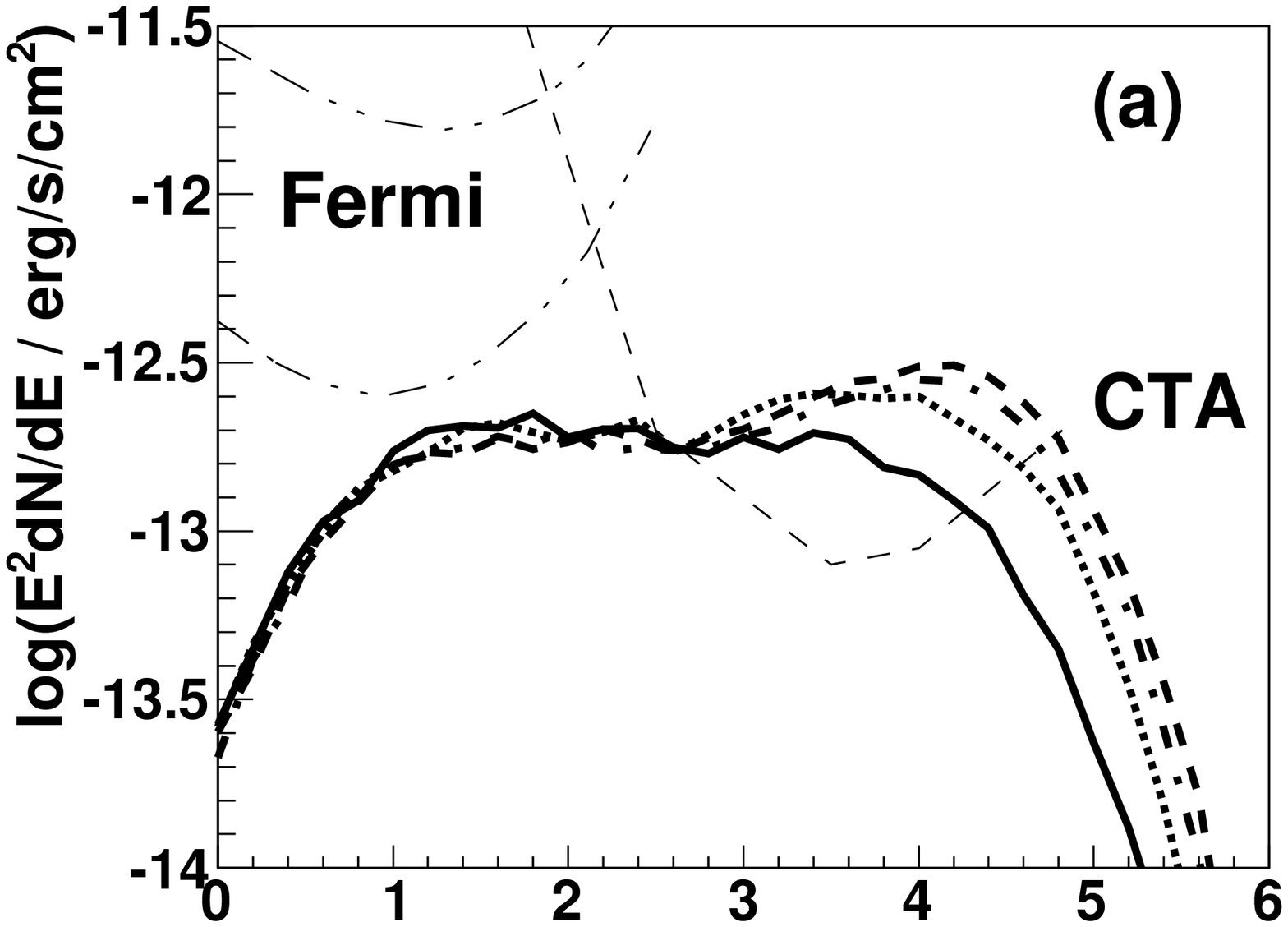}
\includegraphics{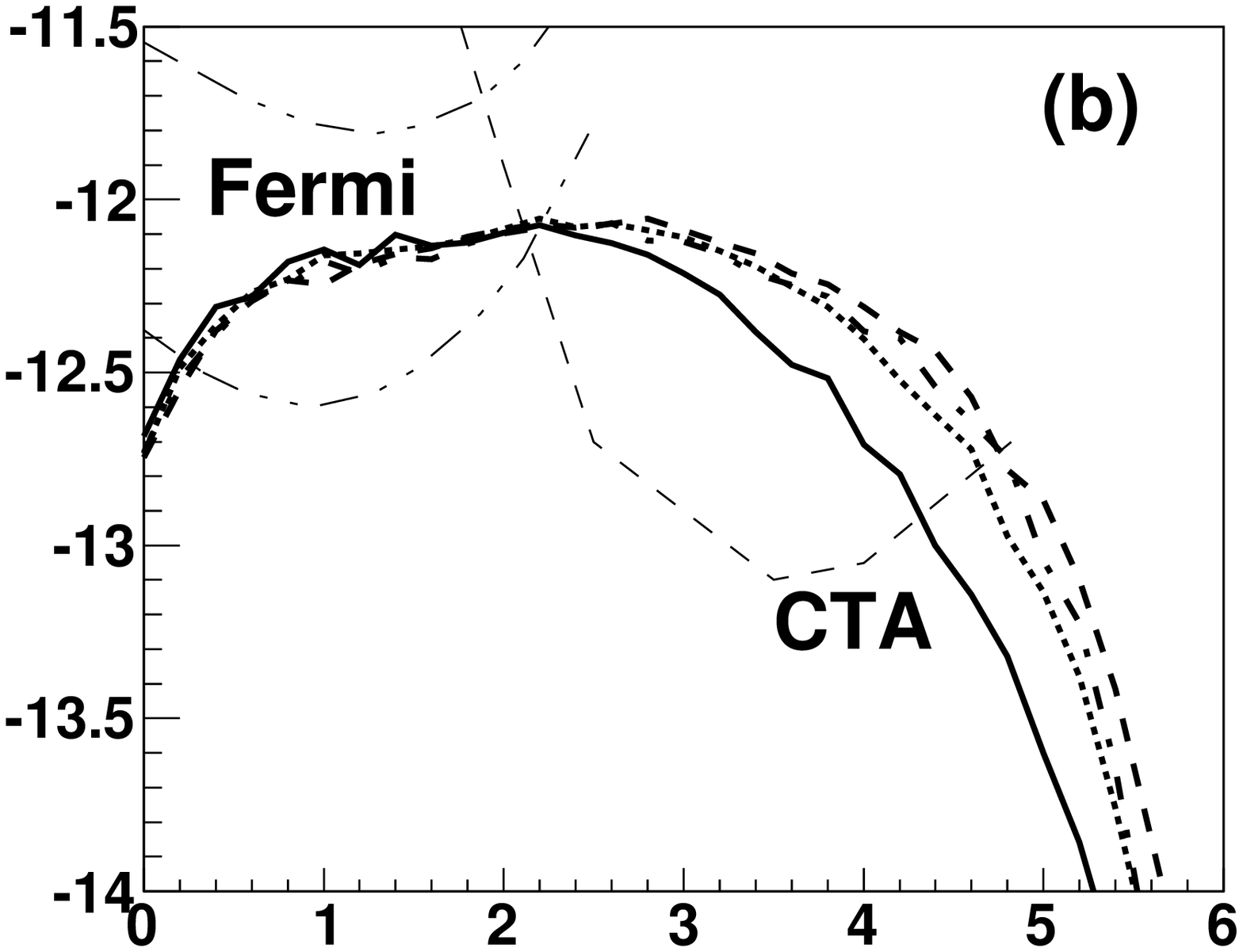}
\includegraphics{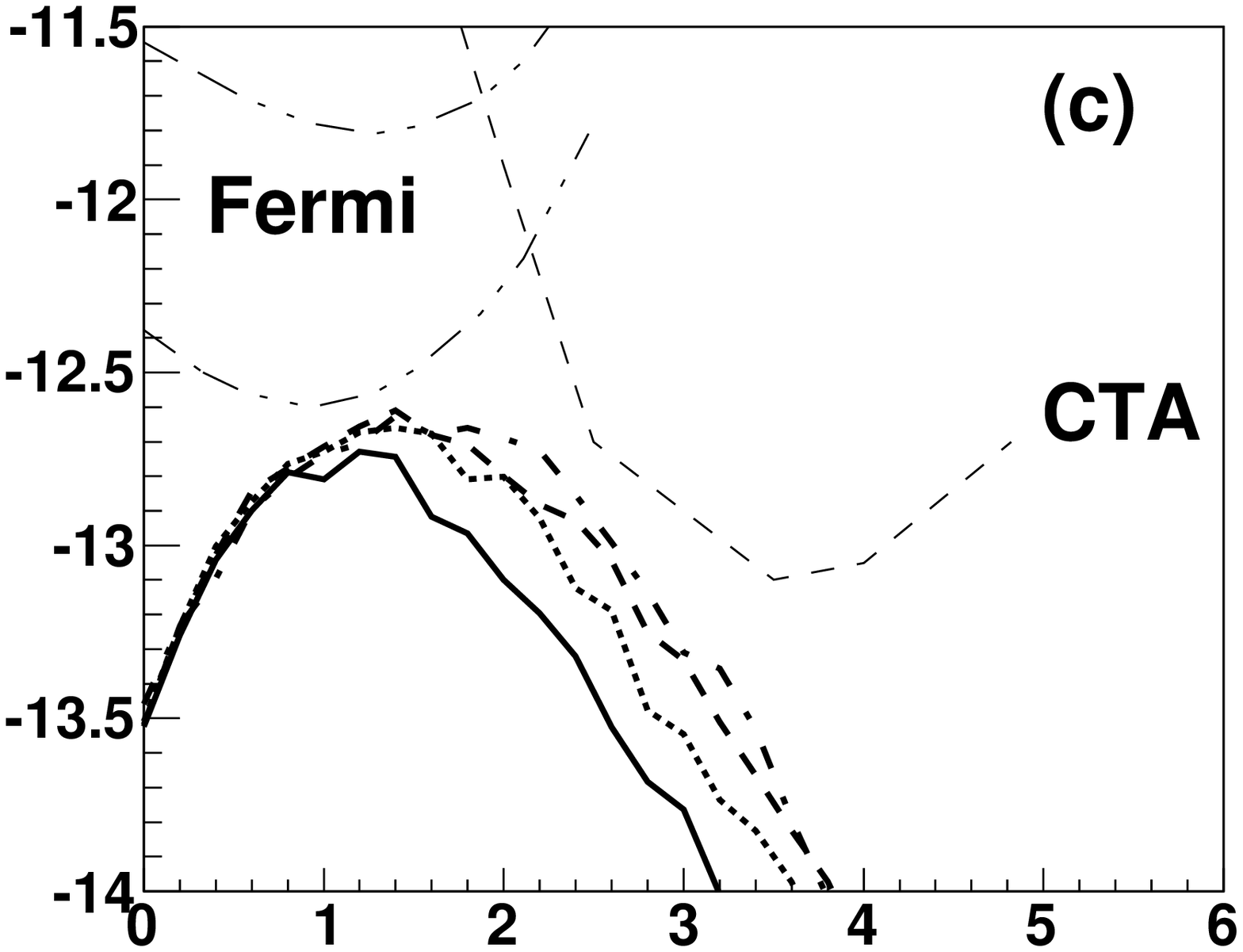}
\includegraphics{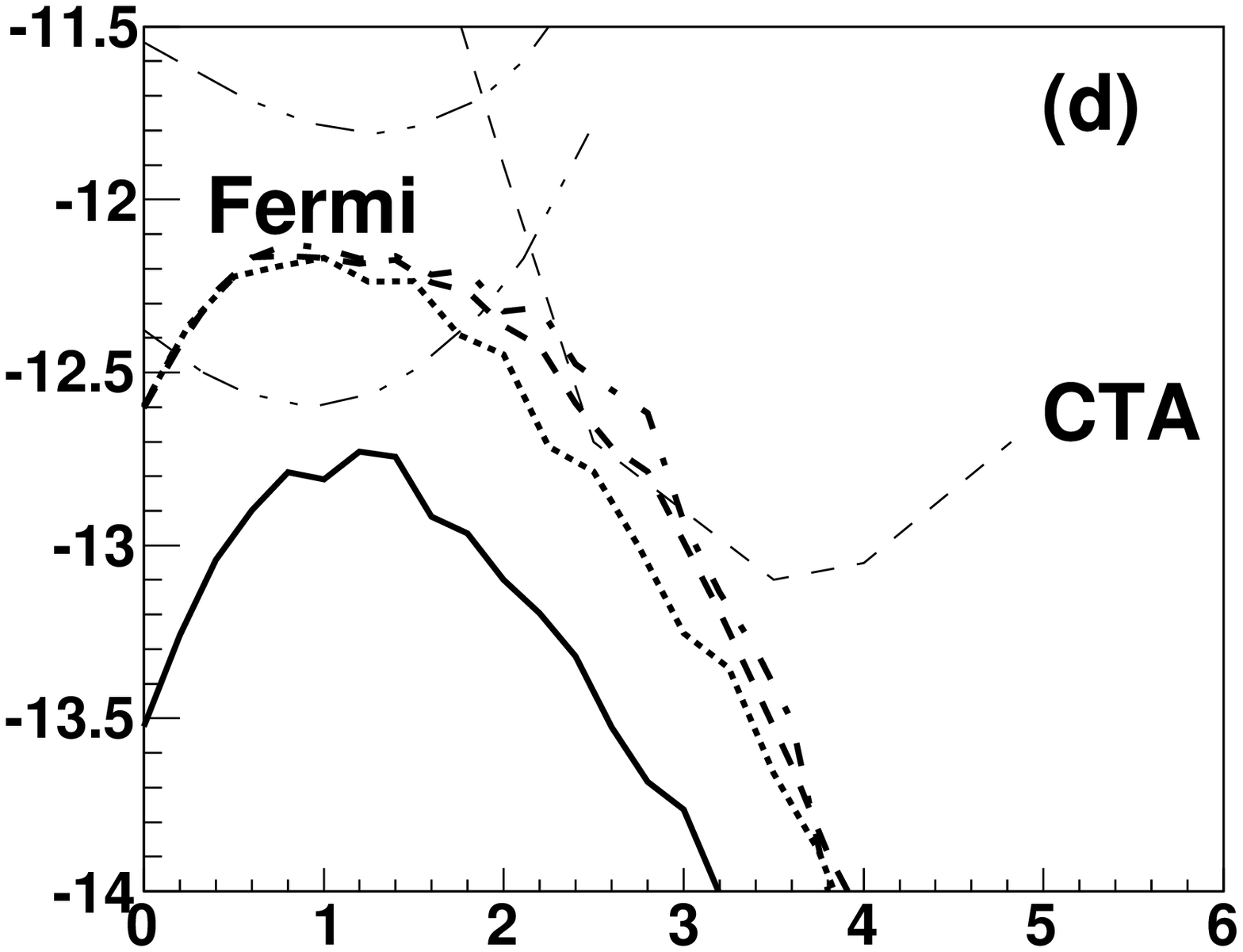}
\includegraphics{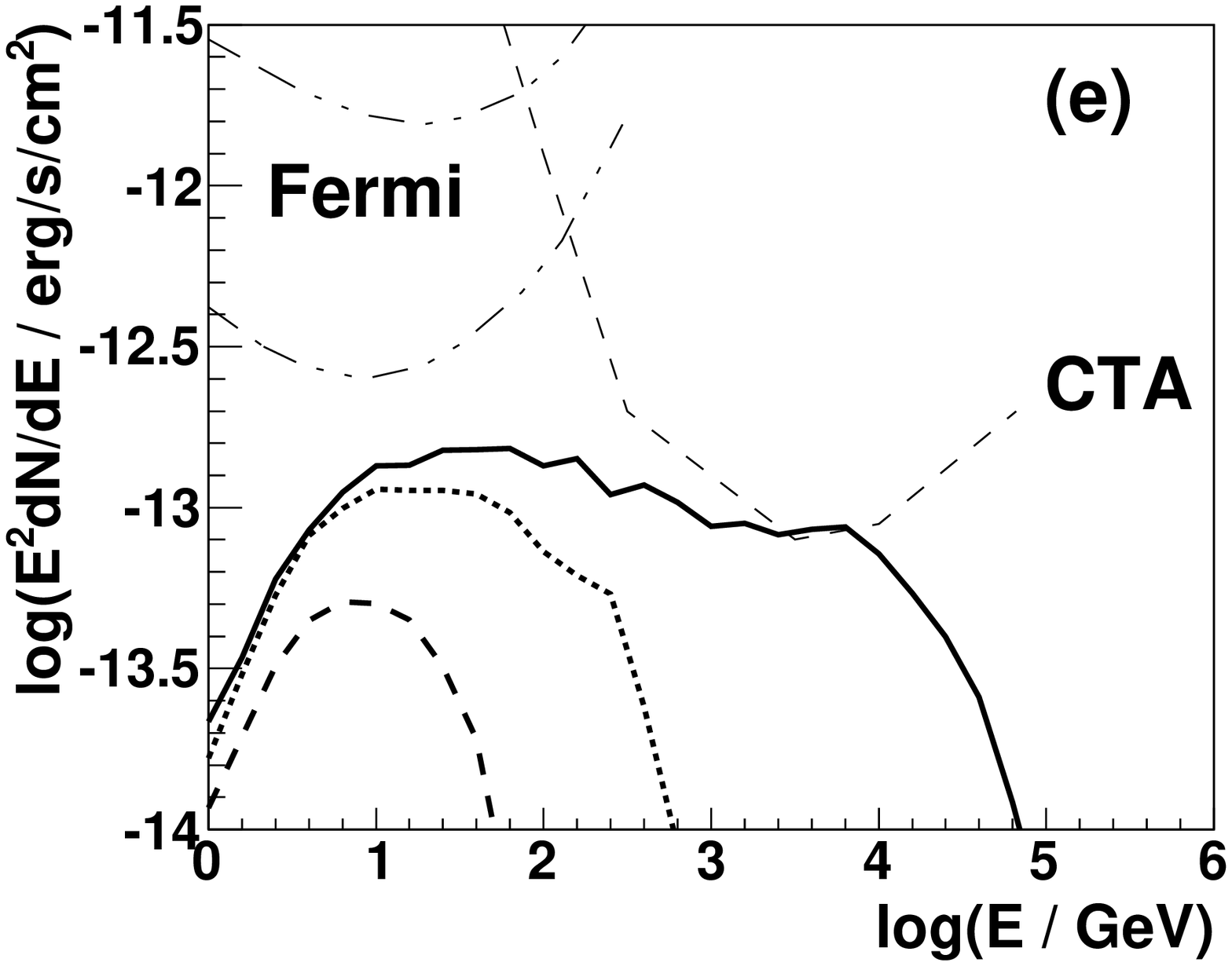}
\includegraphics{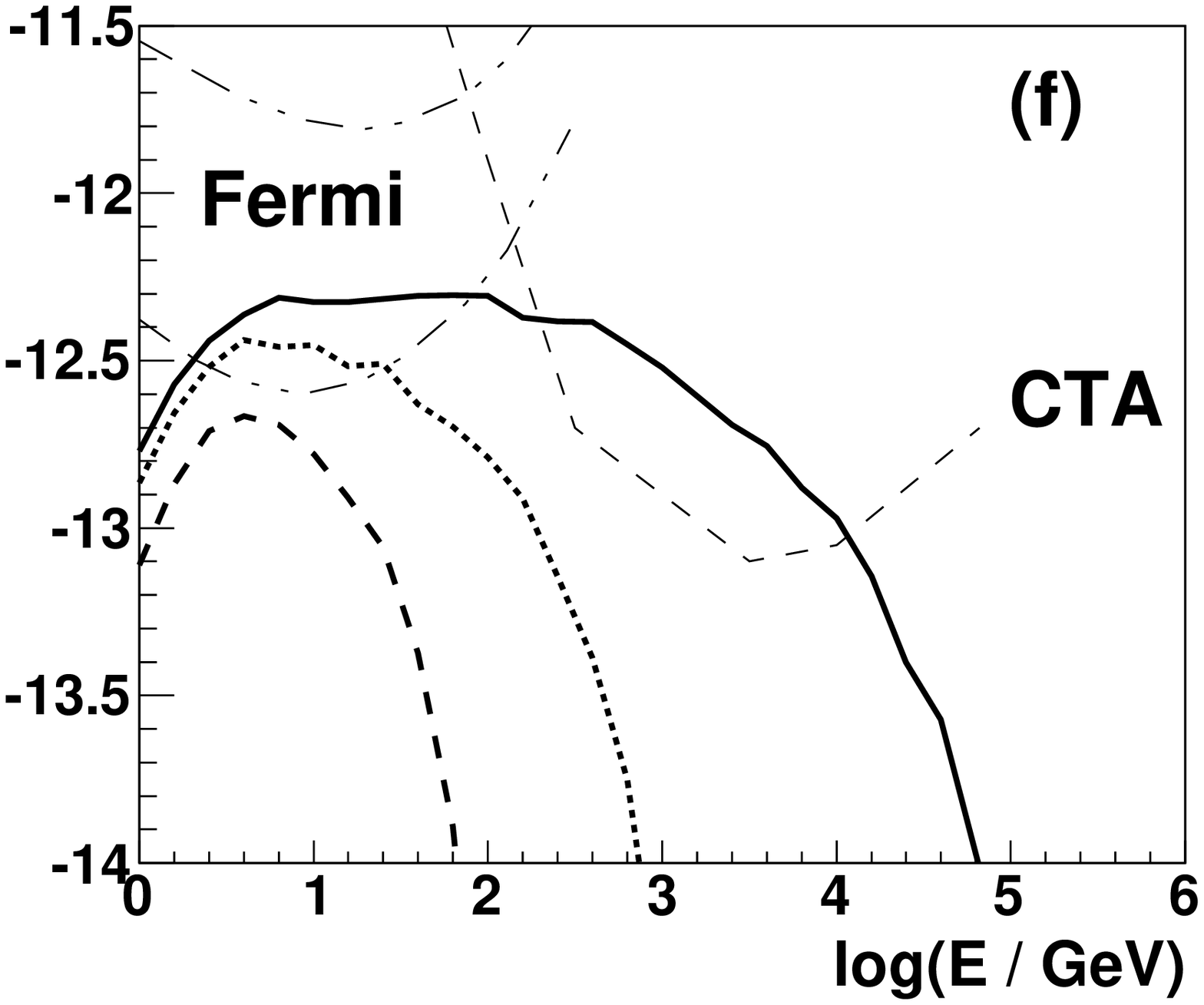}
\includegraphics{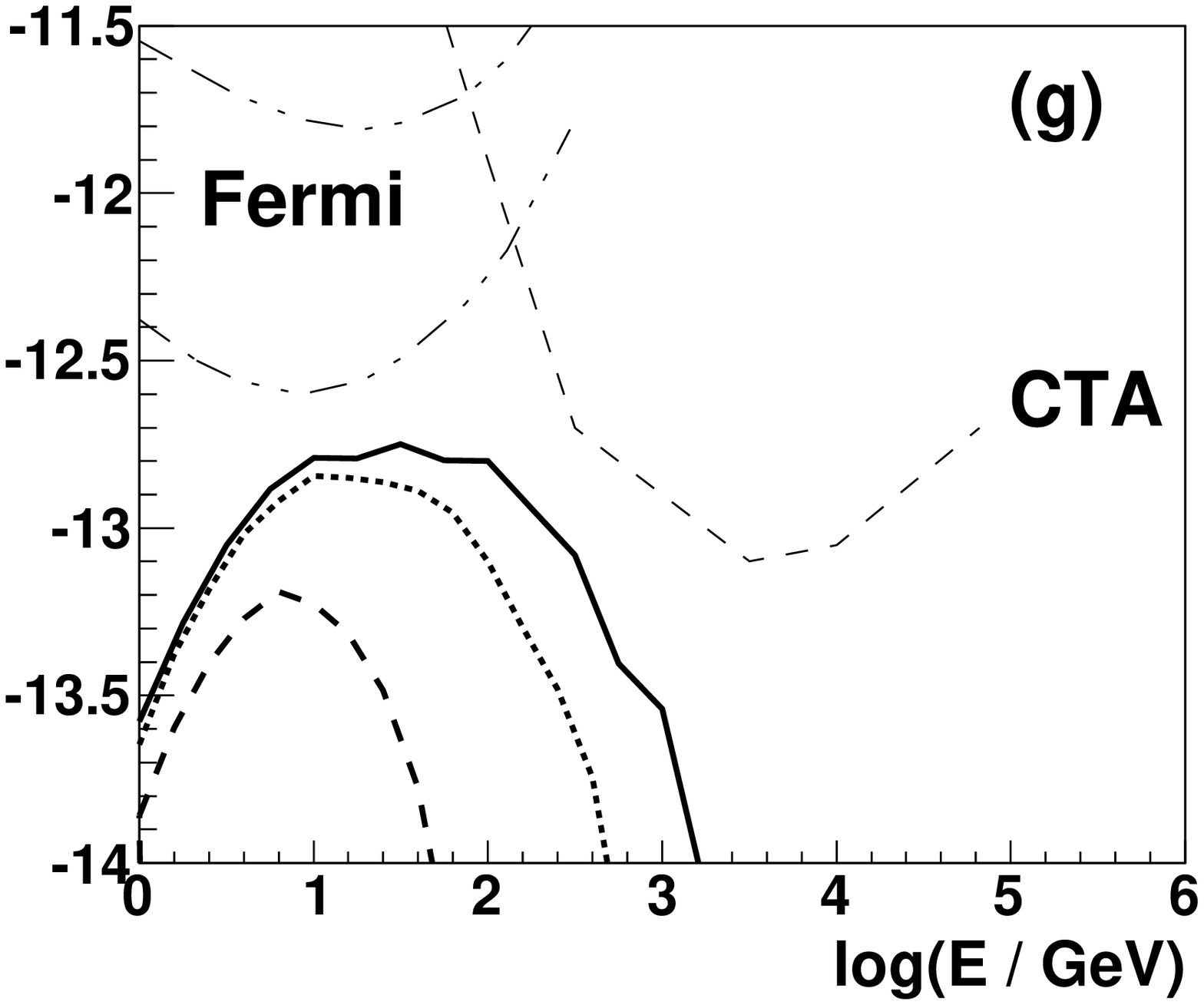}
\includegraphics{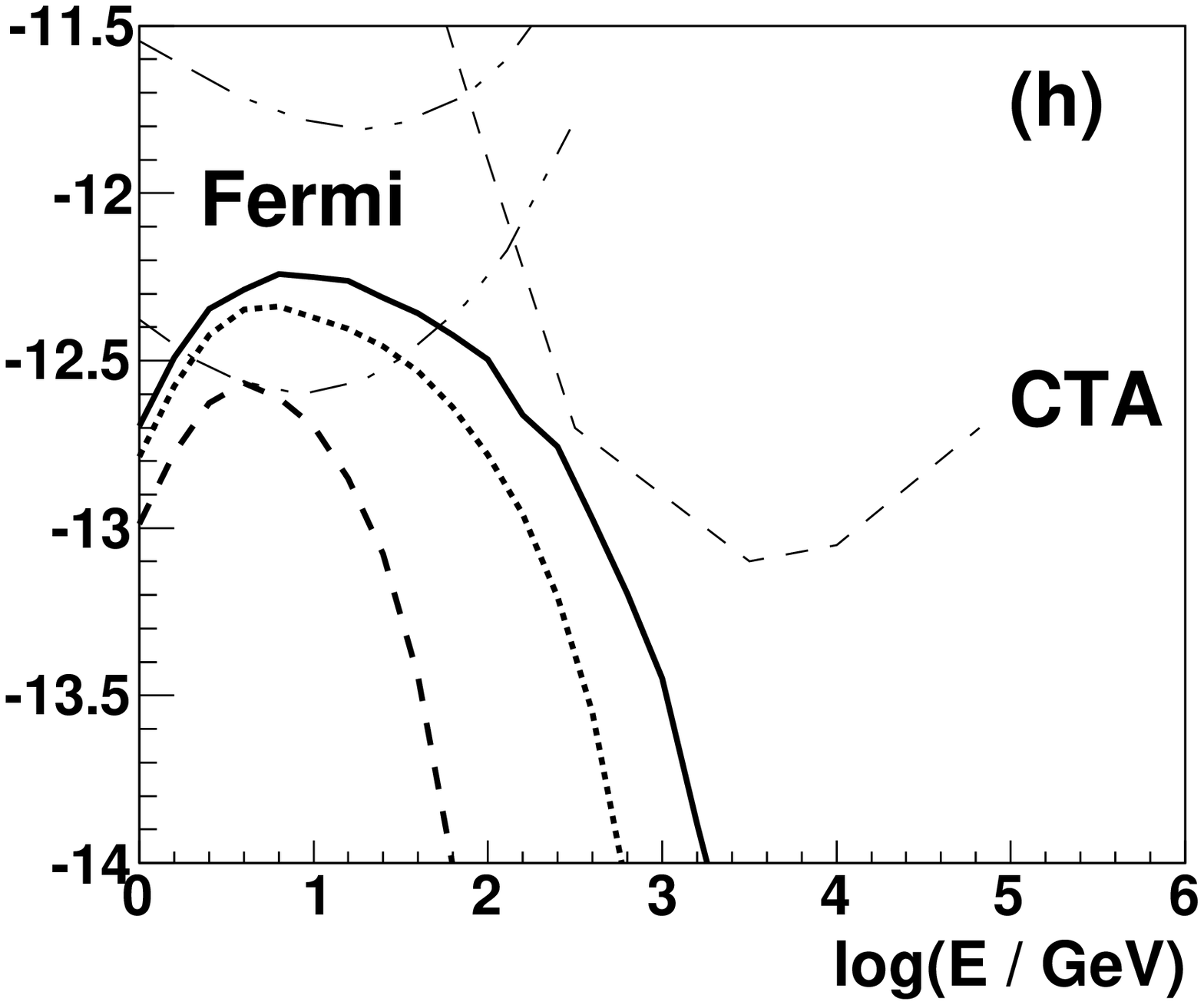}
\caption{The $\gamma$-ray spectral energy distribution  produced in the IC scattering of the RG and the CMBR by relativistic electrons which are accelerated in the polar region of the RG wind. The electrons are accelerated during the initial period after the nova explosion equal to $t_{\rm max} = 15$~yrs (upper panel) and $t_{\rm max} = 30$~days (bottom panel), 
in the sequence of nova explosions with the recurrence time $t_{\rm rec} = 15$~yrs.
They lose energy on the synchrotron process in the NSR magnetic field assumed to be $B_{\rm NN} = 3$~$\mu$G. The electrons were accelerated in the $1-\Omega = 0.7$ fraction of the shells propagating in the polar region of the RG wind (without deceleration),  taking $10\%$ of the kinetic energy of the shells.
The acceleration process of the electrons is defined by the acceleration efficiency $\xi$ and the magnetization parameter of the shell $\alpha$: (a and e) $\gamma$-ray spectra for $\xi = v_{\rm sh}/c$,  $T_{\rm active} = 10^4$~yrs and different magnetization $\alpha = 10^{-5}$ (solid curve), $10^{-4}$ (dotted), $10^{-3}$ (dashed), and $10^{-2}$ (dot-dashed); (b and f) as in (a and e) but for  $T_{\rm active} = 10^5$~yrs. (c, d, g and h)  as in (a, b, e and f) but for the slow acceleration model, i.e. $\xi = 0.1(v_{\rm sh}/c)^2$. The 50 hr sensitivity of the CTA South is marked by the thin dashed curve (see Fig.~2 in 
Maier et al.~2017).
The dot-dot-dashed lines show the 10 yrs \textit{Fermi}-LAT sensitivity \citep{2021ApJS..256...12A}
for Galactic longitude $l=0^\circ$ and latitude  $b=0^\circ$ (upper curve) or $b=30^\circ$ (lower curve)
}
\label{fig5}
\end{figure*}

In the bottom panel of Fig.~5, we show the $\gamma$-ray spectra for these same parameter space as in the upper panel of Fig.~5 but assuming that the electrons
are accelerated in the shells only during their initial stage of propagation after nova explosion, i.e. within the first $t_{\rm max} = 30$ days. This time scale corresponds to the typical period of the GeV 
$\gamma$-ray emission detected from novae (e.g. Ackermann et al.~2014). In this case, only the $\gamma$-ray emission from the nova super-remnant to which the electrons are injected from the weakly magnetized shells with the fast acceleration model are within the sensitivity limit of the CTA (see Fig.~5ef). $\gamma$-rays produced in the slow acceleration model have the chance to be observed only by the satellite telescopes of the \textit{Fermi}-LAT type.

\subsection{Contribution from the equatorial part of the shells}

We also calculate the $\gamma$-ray spectra produced within the nova super-remnant which are expected from the electrons escaping from the ($\Omega$) part of the shells which propagate within the region of the RG wind
 (Fig.~6).
Surprisingly, the shape of these $\gamma$-ray spectra look quite similar to those expected from the electrons escaping from the polar regions of the shells. This is due to two counter-working effects. We argued that the shell in the RG wind decelerates due to entrainment of 
the wind. As a result, the acceleration coefficient of the electrons drops resulting in lower maximum energies of the accelerated electrons. But, from another site, lower velocity of the shell allows more efficient escape 
of electrons from the shell to the NSR. As a result, the effect of decelerating shells becomes compensated.

The $\gamma$-ray spectra from the NSR, produced by the electrons escaping from the decelerating parts of the shells are typically on the level of only a factor of $\sim$2-3 lower than those expected from the polar regions of the shell. For the extreme values of the considered model parameters predicted $\gamma$-ray spectra can be still within the 50 hr sensitivity of the CTA.

\begin{figure*}
\vskip 8.truecm
\includegraphics{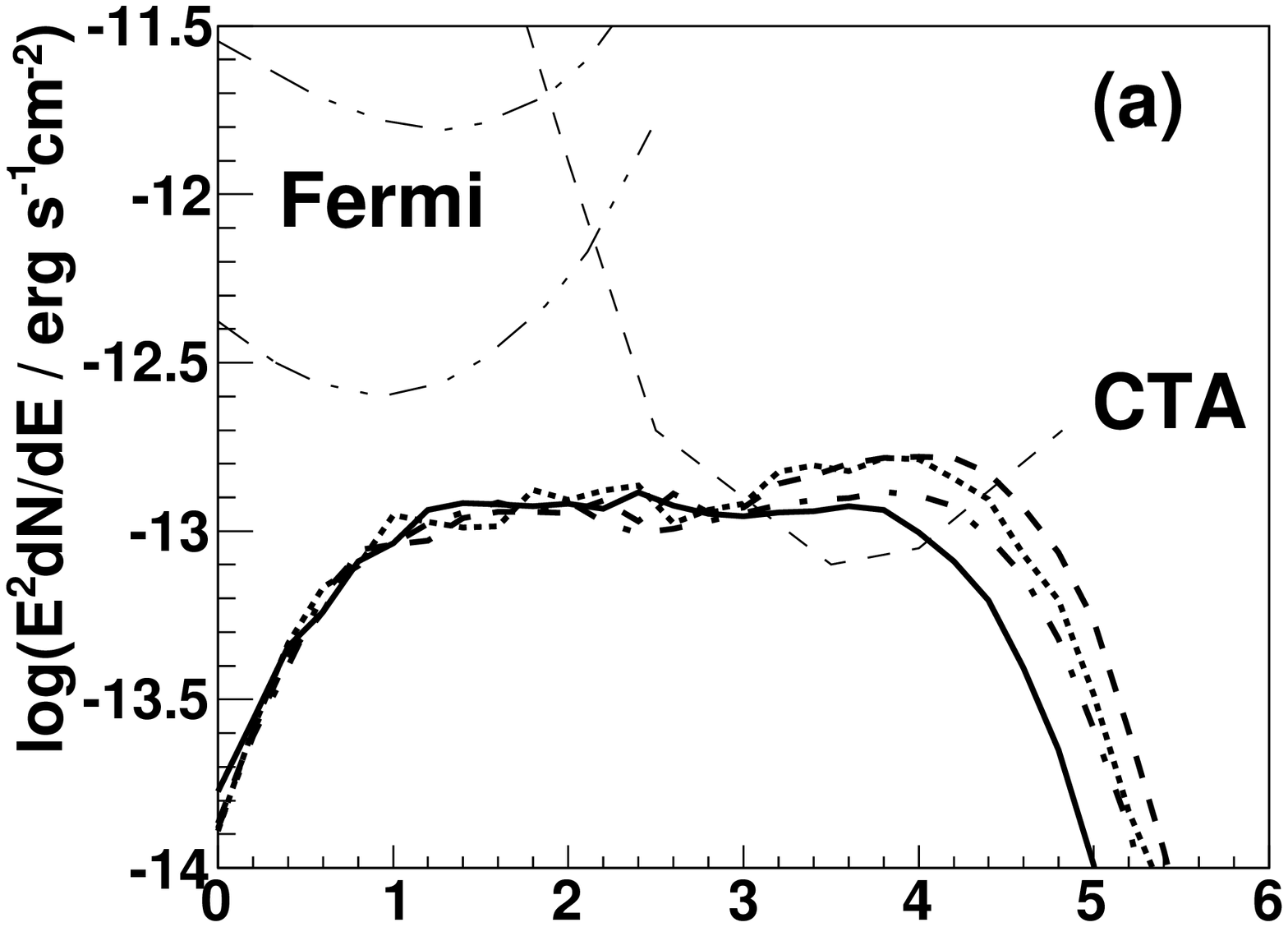}
\includegraphics{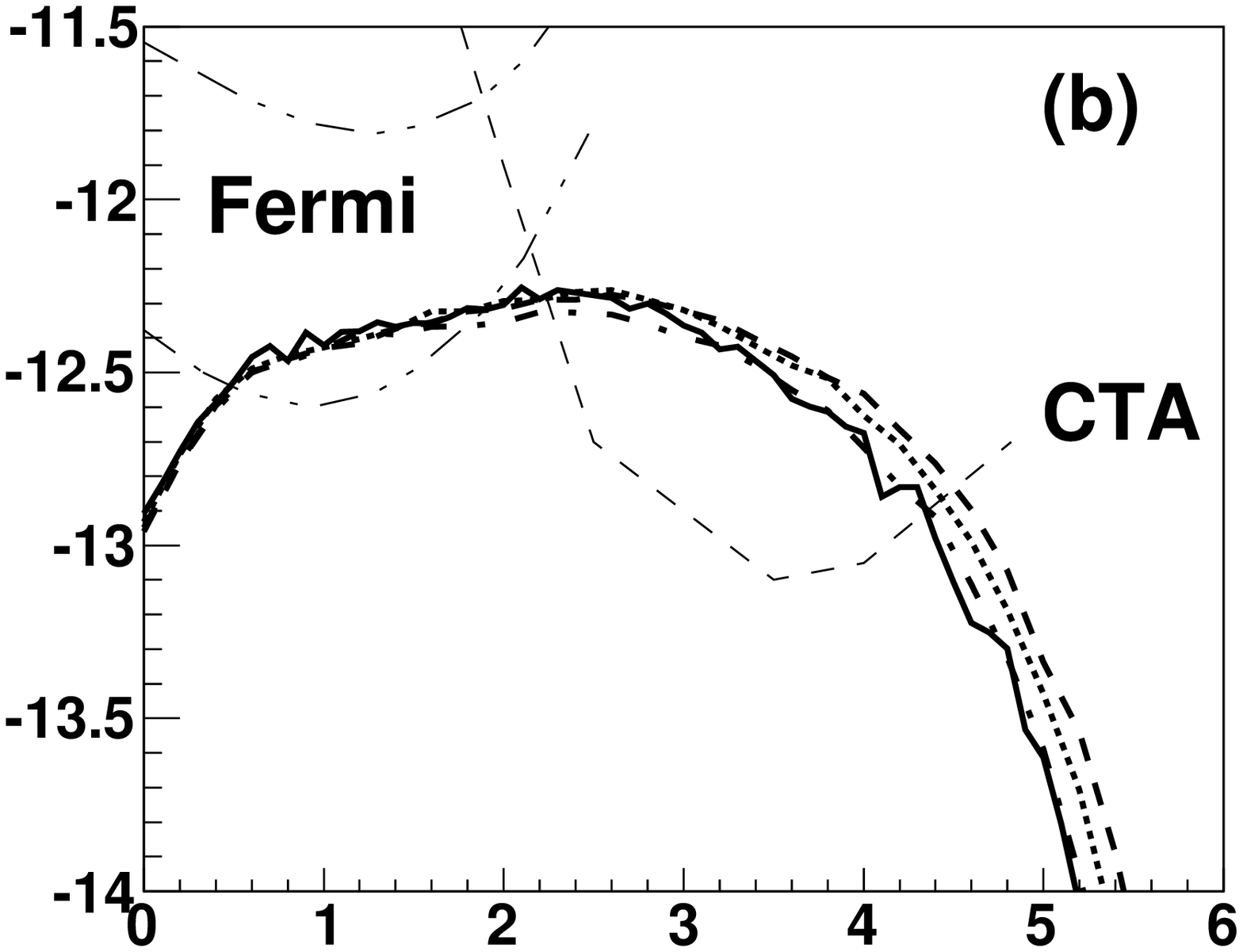}
\includegraphics{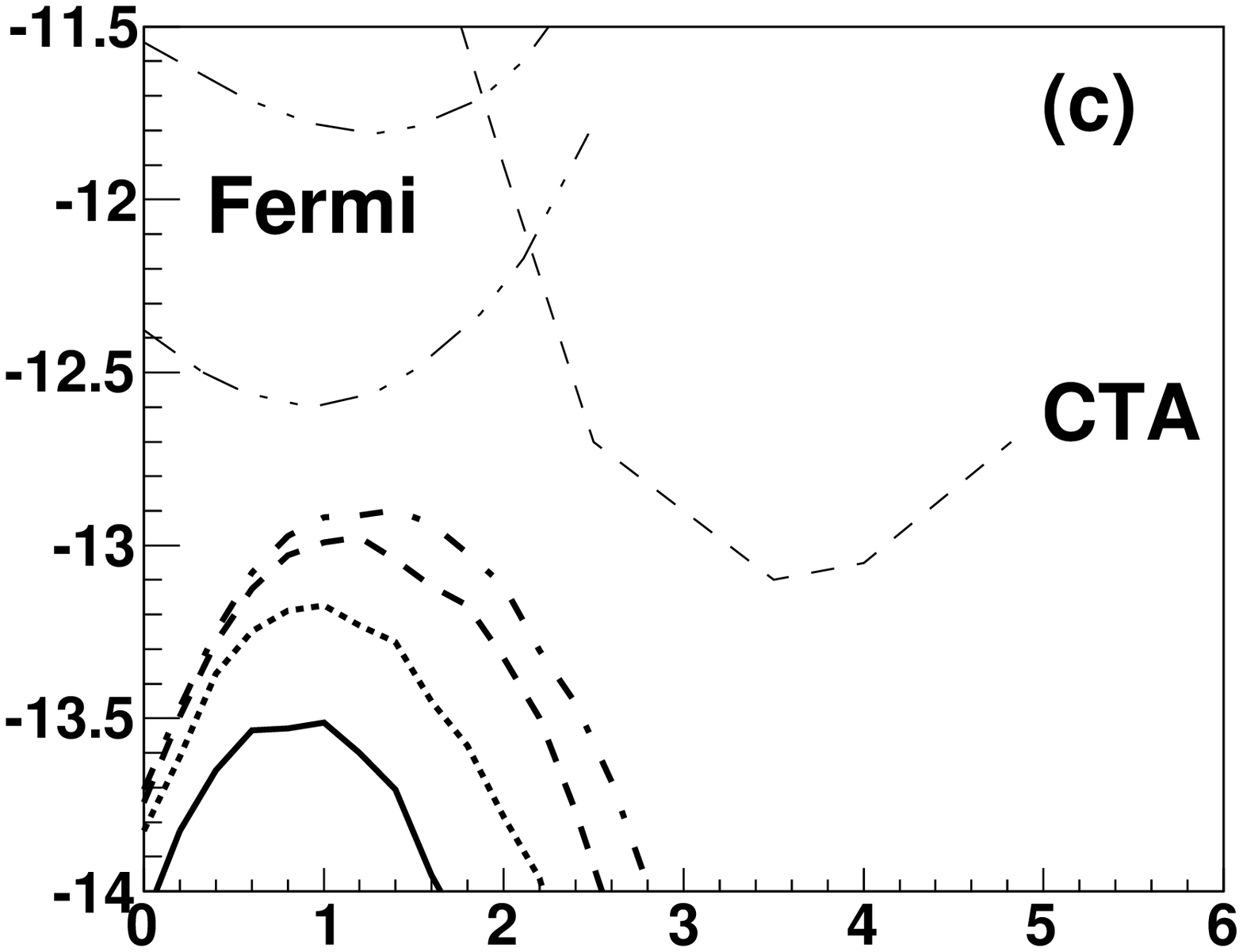}
\includegraphics{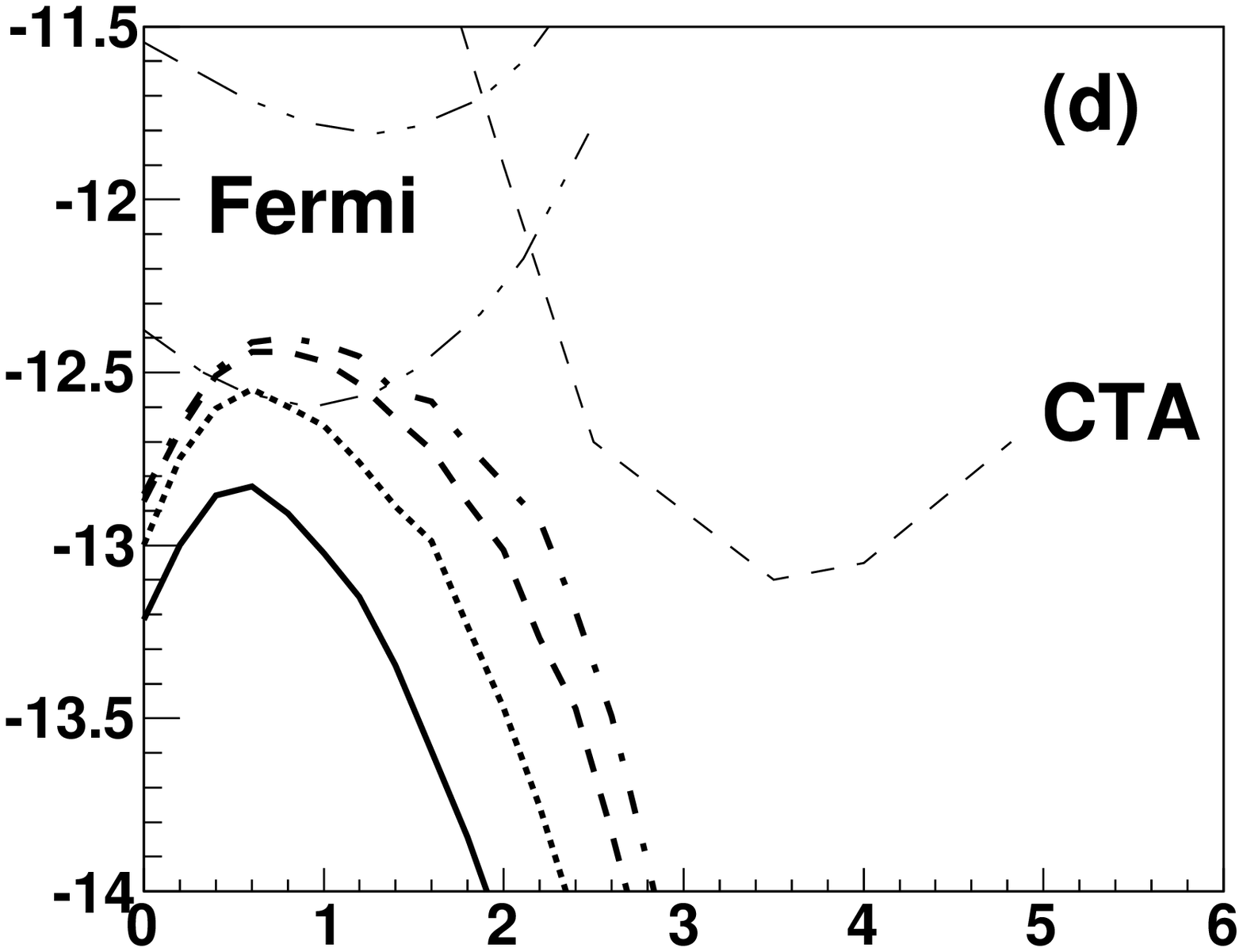}
\includegraphics{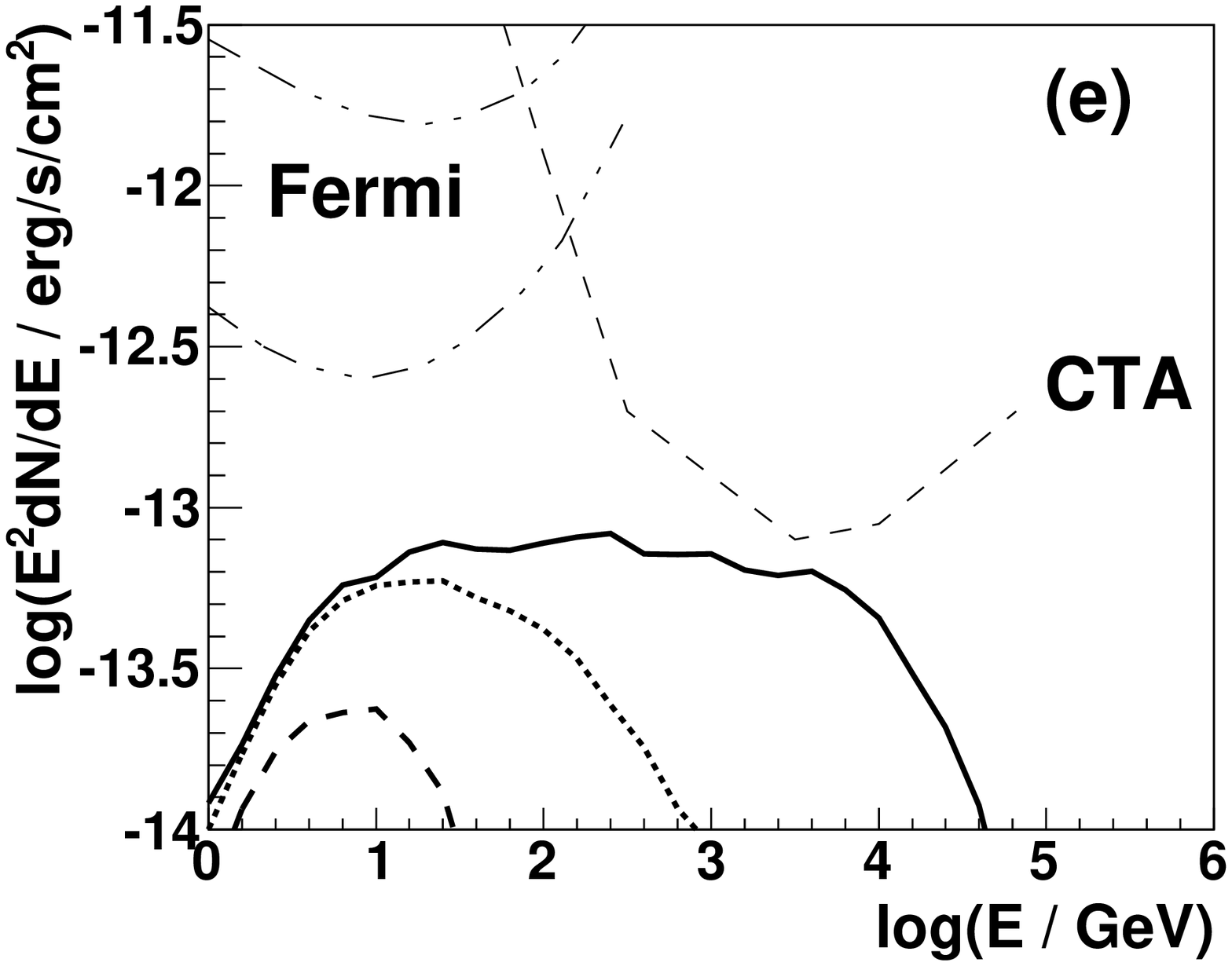}
\includegraphics{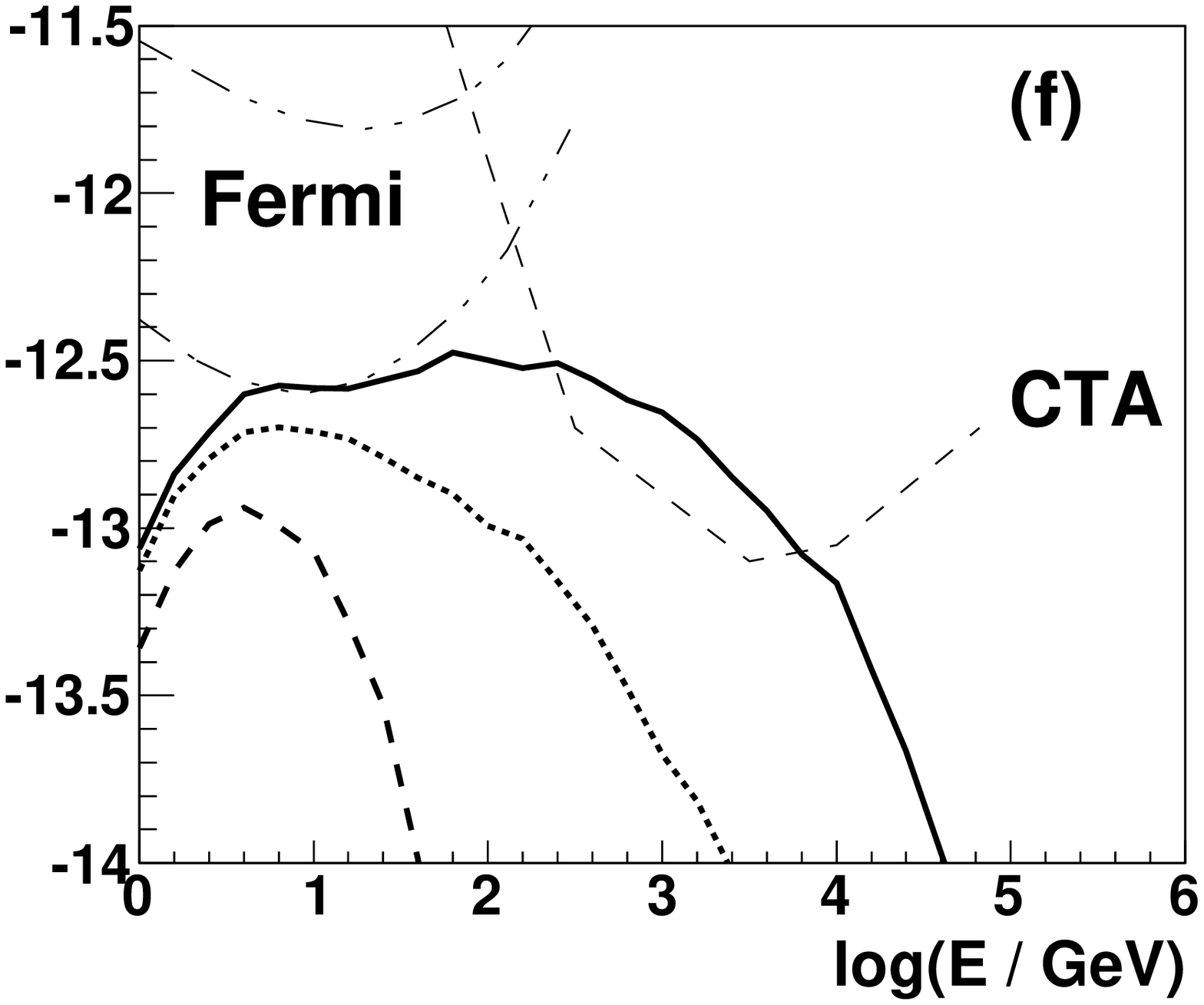}
\includegraphics{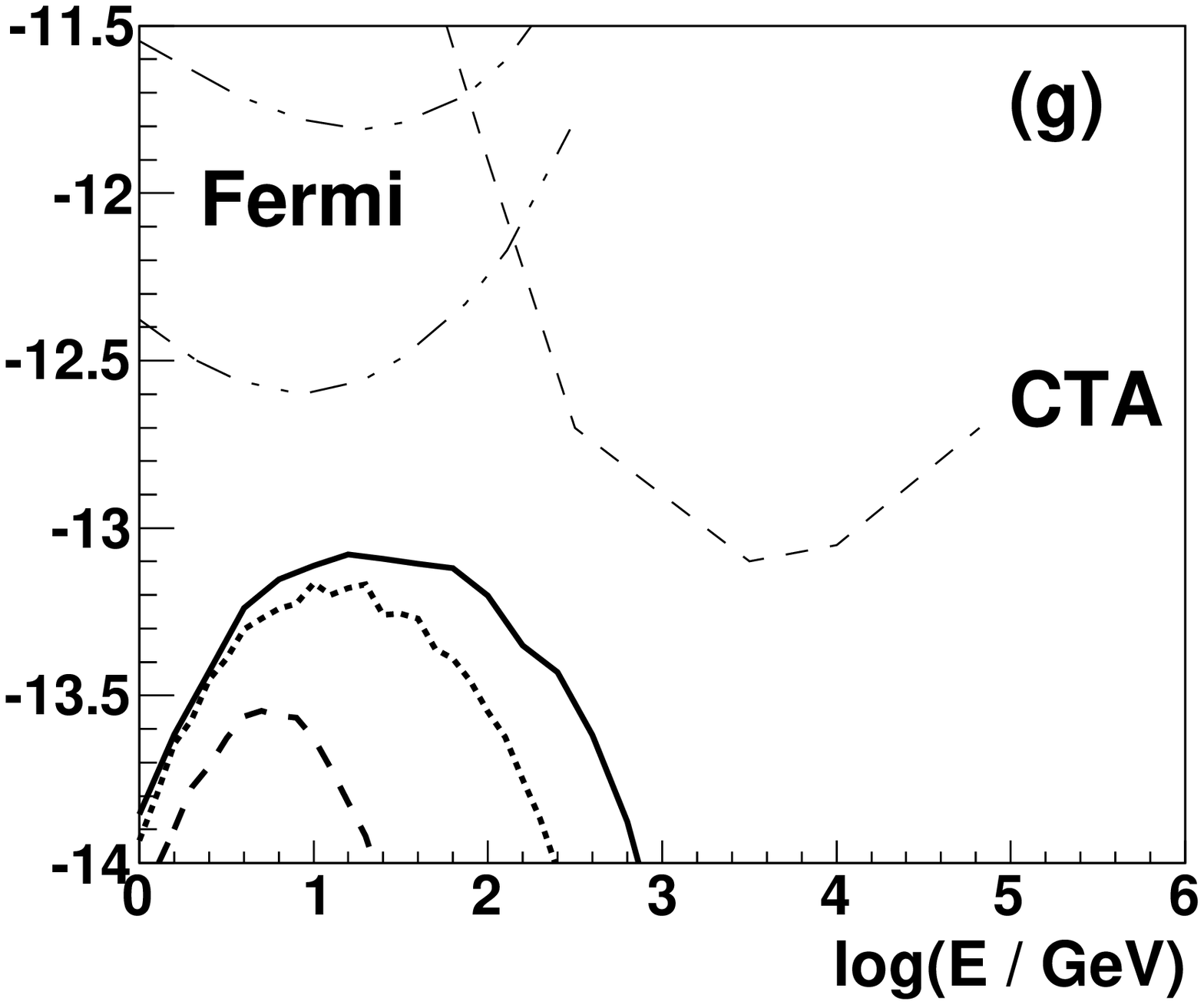}
\includegraphics{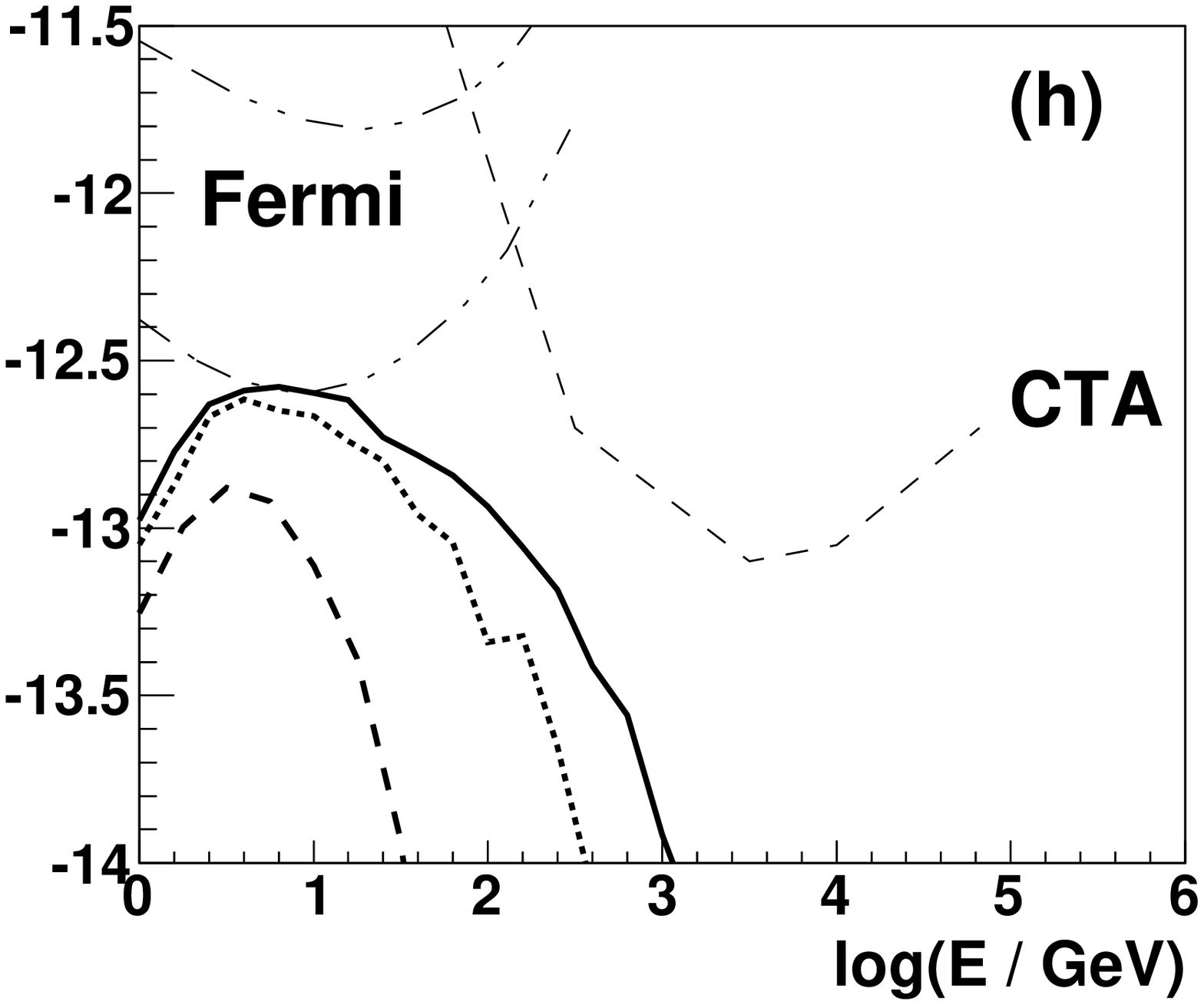}
\caption{As in Fig.~5 but for the part of the shell propagating within the equatorial region of the RG wind with $\Omega = 0.3$. In this case the nova shells are significantly decelerated due to the entrainment of 
the RG wind (see the main text for details).}
\label{fig6}
\end{figure*}
\subsection{Dependence of gamma-ray emission on other model parameters}

We also study the effect of other model parameters on the $\gamma$-ray emission from NSR. 
All the calculations of the $\gamma$-ray emission from the NSRs have been performed assuming that the relativistic electrons escape from  the shells of novae at distances not larger than $R_{\rm inj} = 2\times 10^{17}$~cm. $R_{\rm inj}$ corresponds to the characteristic distance at which nova shells start to significantly decelerate due to interaction with the interstellar space. Its value depends on the local density of interstellar space.  
In Fig.~7b, we investigate the effect of $R_{\rm inj}$ on the $\gamma$-ray spectrum from NSR. We show the spectra for $R_{\rm inj} = 10^{16}$~cm -- $10^{18}$~cm. 
We conclude that the $\gamma$-ray spectra do not depend significantly on the injection distance 
$R_{\rm inj}$ at the TeV energies. However, lower energy electrons, injected at large distances from the nova binary system, are not able to cool efficiently in a already weak radiation field of the RG companion star. Their energy losses on the CMBR are also inefficient.

Finally, in Fig.~7c, we investigate the dependence of the $\gamma$-ray spectrum on the activity period of the NSR, T$_{\rm active}$.  The $\gamma$-ray emission is on a comparable level at multi-TeV energies, independent on the age of the NSR for the considered range of ages, i.e. $10^4-10^7$~yrs. In fact, the collision time of the electrons with the CMBR can be estimated on 
\begin{eqnarray}
\tau_{\rm CMBR}\sim (n_{\rm CMBR}\sigma_{\rm T}c)^{-1}\approx 4.4\times 10^3~~~{\rm yrs}, 
\label{eq1}
\end{eqnarray} 
\noindent
where $n_{\rm CMBR}$ is the density of the CMBR and $\sigma_{\rm T}$ is the Thomson cross section. 
$\tau_{\rm CMBR}$ is clearly shorter than considered by us activity period of the recurrent Nova.
The scattering process of the CMBR contributes mainly to the $\gamma$-ray spectrum at GeV energies.
This part of the $\gamma$-ray spectra saturate for the large activity time of the recurrent nova
(see dashed and dot-dashed curves in Fig.~7c).  
We conclude that the high energy electrons in the NSR can lose efficiently energy during 
the time $T_{\rm active}$.

\begin{figure*}
\vskip 5.5truecm
\includegraphics{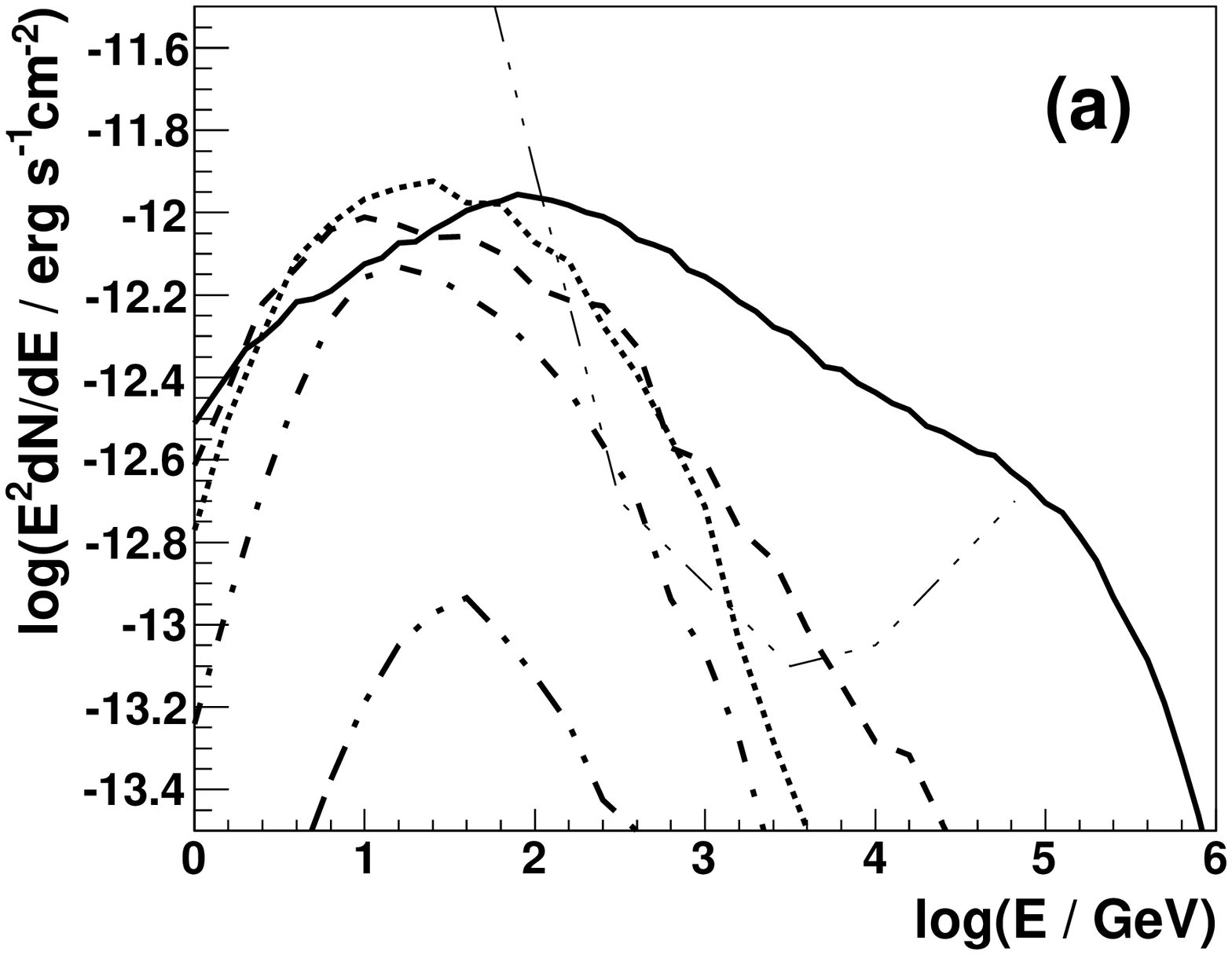}
\includegraphics{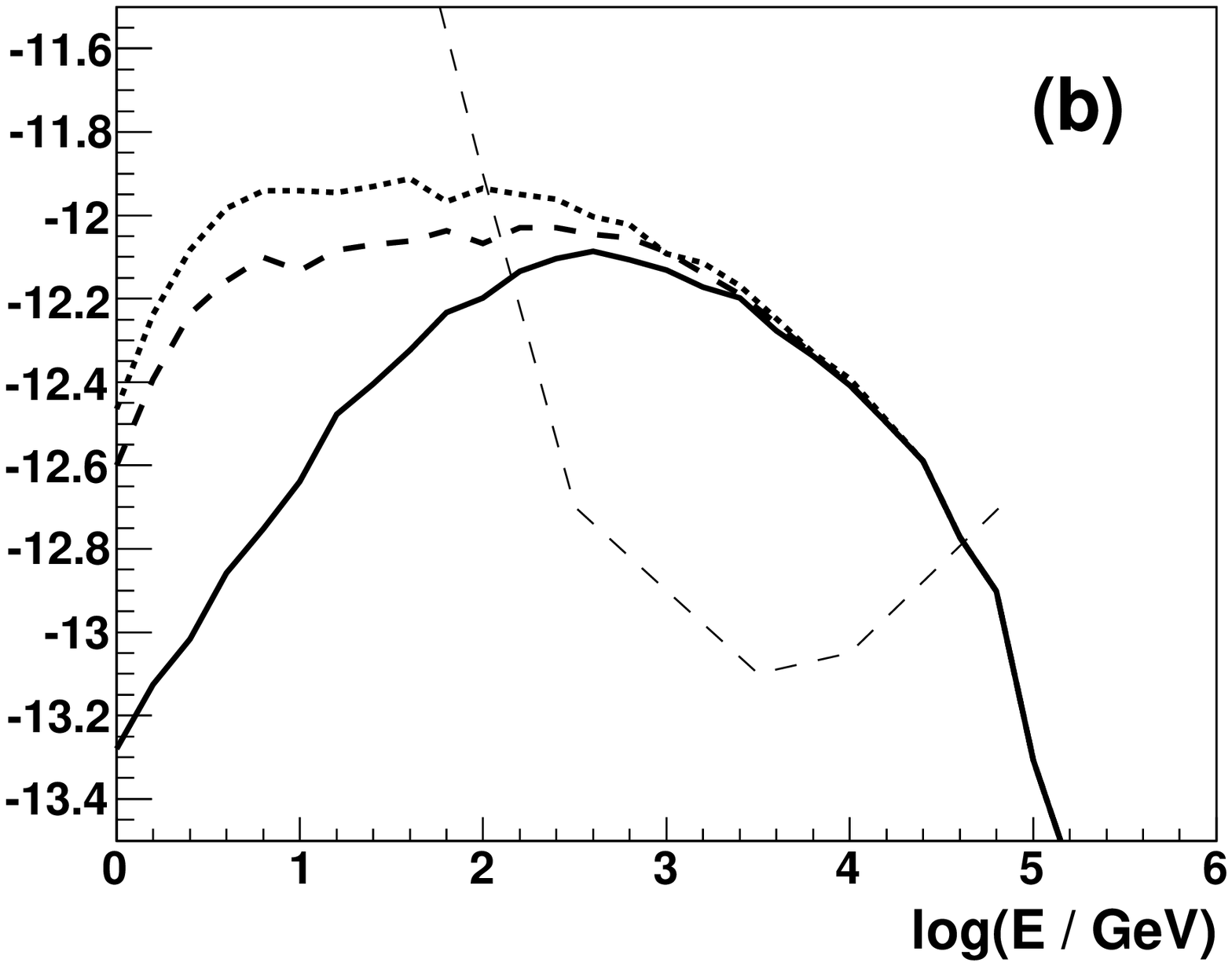}
\includegraphics{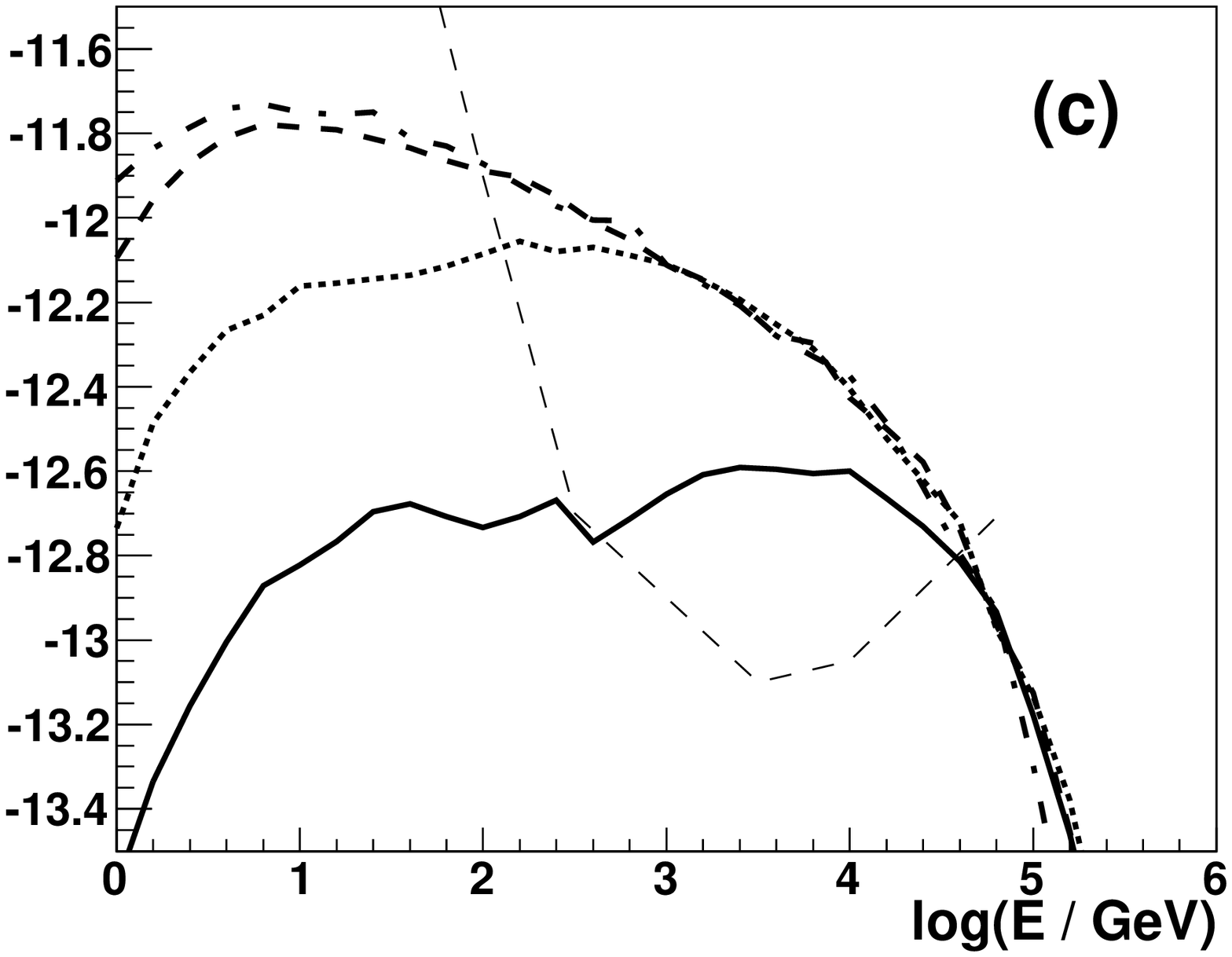}
\caption{The $\gamma$-ray spectral energy distribution as a function of: (a) the strength of the magnetic field within the NSR $B_{\rm NSR} =3$~$\mu$G (solid), 10~$\mu$G (dashed), 30~$\mu$G (dotted), 100~$\mu$G (dot-dashed), and 300~$\mu$G (dot-dot-dashed) (a). The other parameters are $\alpha = 10^{-4}$,  $R_{\rm inj} = 2\times 10^{17}$~cm,
$T_{\rm rec} = 15$~yrs,  $T_{\rm active} = 10^5$~yrs, and $\xi = 0.01$; (b) the injection distance of relativistic electrons within the NSR $R_{\rm inj} = 10^{16}$~cm (dotted curve), $10^{17}$~cm (dashed), and $10^{18}$~cm (solid). The other parameters are as above and $B_{\rm NSR} = 3$~$\mu$G; and (c) as a function of the activity stage of the nova $T_{\rm active} = 10^4$~yrs (solid), $10^5$~yrs (dotted), $10^6$~yrs (dashed), and $10^7$~yrs (dot-dashed). The other parameters are the same as in (a) and (b).}
\label{fig7}
\end{figure*}

\subsection{Synchrotron emission from electrons in NSR}

All the above calculations have been performed under assumption that the magnetic field strength is of the order of that observed in the interstellar space (i.e. $B_{\rm NSR} = 3\mu$G). In principle, the magnetic field in the surrounding of the recurrent nova might build up to larger values as observed in the pulsar wind nebulae. 
For example, this stronger magnetic field can be provided to the inner parts of the NSR with the multiple nova shells. 
In Fig.~7a, we show the $\gamma$-ray spectra calculated for much stronger values of the magnetic field 
$B_{\rm NSR}$. As expected, the TeV part of the $\gamma$-ray spectrum is strongly suppressed for stronger magnetic fields due to the efficient synchrotron energy losses of the electrons. However, the GeV $\gamma$-ray emission is still on a comparable level provided that $B_{\rm NSR} < 100\mu$G.  For very strong magnetic fields, i.e. $B_{\rm NSR} > 100\mu$G, the GeV $\gamma$-ray emission becomes also strongly suppressed.
Therefore, we conclude that strongly magnetised NSRs around novae have still a chance to be detected at
GeV $\gamma$-ray energies but not at TeV energies. 

The TeV $\gamma$-ray emission from NSR is expected to be very sensitive on the strength of the magnetic field within the NSR. If $B_{\rm NN}$ is larger than 100~$\mu$G, then the TeV 
$\gamma$-ray emission is below sensitivity of the CTA even for the most favourite parameters of the model
(see Fig.~7a). 
However, the $\gamma$-ray emission in the GeV energy range is quite stable, provided that
$B_{\rm NSR} < 100$~$\mu$G. If the magnetic field within NSR is significantly stronger than 3~$\mu$G, then the synchrotron emission from TeV electrons is expected in the range from the infrared to the X-rays,
$\varepsilon_{\rm syn} = m_{\rm e}c^2(B/B_{\rm cr})\gamma_{\rm e}^2\sim 0.05B_{\mu G}E_{\rm TeV}^2$~eV.
We show the expected synchrotron spectra for the example parameters of the model (as used in Fig.~7a), as a function of the magnetic field within the NSR in Fig.~8. As expected the synchrotron emission starts to dominate over the $\gamma$-ray IC emission from the NSR for stronger magnetic fields.
We note that this synchrotron emission in the optical range is a few orders of magnitudes below the optical
emission from the RG. 
The thermal X-ray emission, detected in quiescence from RS Oph (e.g. Nelson et al.~2011) or the extended emission observed in the polar regions of the RS Oph during the outburst in 2006 (Montez et al.~2022)
comes from a relatively very small region (of the order of arc seconds) around the nova RS Oph. 
This emission cannot constrain synchrotron X-ray emission expected in our model which comes from much more extended $\gamma$-ray NSR (see Eq.~7).

\begin{figure}
\vskip 5.2truecm
\includegraphics{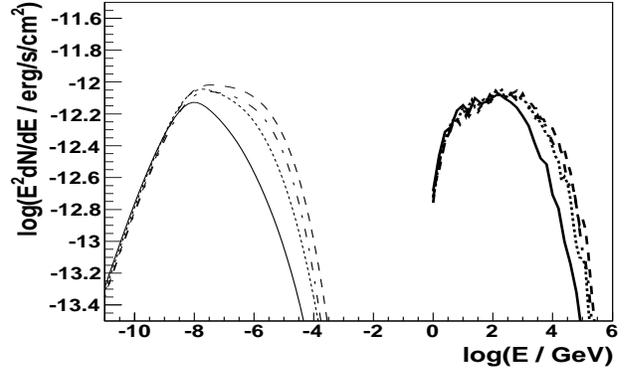}
\caption{The IC (thick curves) and the synchrotron (thin) spectra calculated for different strength  of the magnetic field within the NSR expected around RS Oph: $B_{\rm NSR} =3$~$\mu$G (solid), 10~$\mu$G (dashed), 30~$\mu$G (dotted) and 100~$\mu$G (dot-dashed). The other parameters are as in Fig.~7a. }
\label{fig8}
\end{figure}
\section{Estimate of the nova super-remnant size}

The propagation process of charged particles in the vicinity of compact sources is not clear. Therefore,
we cannot definitely conclude on the size of the NSR. In the calculations above, we applied the Bohm diffusion prescription. But in fact, two limiting scenarios can be considered.
If the magnetic field around  the nova is ordered (e.g. magnetic field mainly perpendicular to the radial direction from the binary system), then $\gamma$-rays produced in the IC process can be arbitrarily confined 
forming a point like source. On the contrary, the magnetic field can be random, resulting in a rapid diffusion process, e.g. consistent with the Bohm diffusion prescription. In this second limiting case, we can simply estimate the maximum size of the NSR provided that injected electrons lose energy on the IC process by scattering CMBR and RG radiation.

We estimate the size of the $\gamma$-ray region from the NSR for the electrons
with energies $E_\gamma = 1E_{\rm TeV}$ TeV. Such electrons produce $\gamma$-ray photons with typical energies within sensitivity of the {\it Fermi}-LAT telescope,
$E_\gamma\sim \varepsilon_{\rm CMBR}\gamma_{\rm e}^2\sim 2.8E_{\rm TeV}^2$~GeV as a result of comptonization of the CMBR, where $\varepsilon_{\rm CMBR}\sim 7\times 10^{-4}$~eV. They also produce $\gamma$-rays by scattering the RG radiation with energies comparable to the energy of the electrons, i.e. typical 
for the Cherenkov telescopes.
The cooling time scale of electrons (in the Thomson regime) with energy equal to 1 TeV in the CMBR is 
\begin{eqnarray}
\tau_{\rm IC}^{\rm CMBR}\approx 3.2\times 10^{13}/E_{\rm TeV}~~~{\rm s}.
\label{eq1}
\end{eqnarray} 
\noindent
and in the RG radiation
\begin{eqnarray}
\tau_{\rm IC}^{\rm RG}\approx 3\times 10^{10}R_{17}^2/E_{\rm TeV}~~~{\rm s}.
\label{eq1}
\end{eqnarray} 
\noindent
where $R = 10^{17}R_{17}$ cm is the distance measured from the RG, that determines the radiation field.

Bohm diffusion distance of these electrons during the cooling time scale on the CMBR, $t = \tau_{\rm IC}^{\rm CMBR}$, is
\begin{eqnarray}
x_{\rm dif, CMBR} = \sqrt{2 D_{\rm B} t}\approx 29/B_{\rm NSR,\mu G}^{1/2}~~~{\rm pc},
\label{eq2}
\end{eqnarray} 
\noindent
where $B_{\rm NSR,\mu G}$ is $B_{\rm NSR}$ in units of $\mu G$. 
During the cooling time scale on the RG radiation the nominal diffusion distance is 
\begin{eqnarray}
x_{\rm dif, RG}\approx 0.3R_{17}/B_{\rm NSR,\mu G}^{1/2}~~~{\rm pc},
\label{eq:xdifRG}
\end{eqnarray} 
\noindent
Note that while $x_{\rm dif, RG} < x_{\rm dif, CMBR}$ would suggest the dominant role of the RG radiation field in the cooling and limiting the extend of the diffusion this is not the case.
For a broad range of parameters ($B_{\rm NSR,\mu G} < 100\mu G$), Eq.~\ref{eq:xdifRG} states that 
$x_{\rm dif, RG} > R = 10^{17}R_{17}$~cm. Therefore as the electrons diffuse they experience steeply falling radiation field from the RG (as $R^{-2}$), that is not able to cool them down. 
We conclude that the IC cooling of electrons on the RG radiation do not constrain the size of the NSR.

\begin{figure}
  \includegraphics[width=0.49\textwidth]{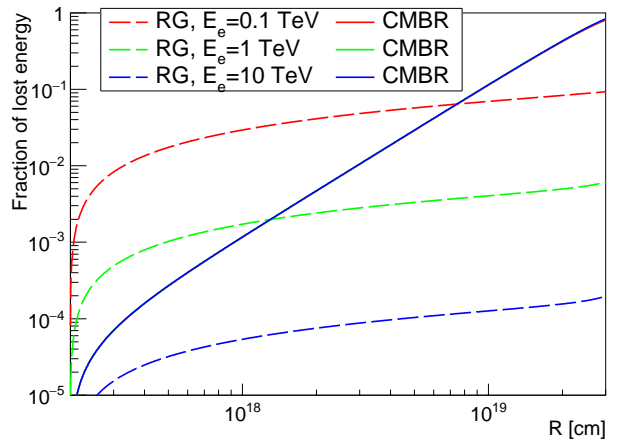}
  \caption{
Energy loss of the electrons on the radiation field of RG (dashed lines) and on CMBR (solid lines, merging together) in their diffusion through the magnetic field of 3$\mu$G. Electrons are injected at the distance of $2\times 10^{17}$~cm with energies 0.1~TeV (red), 1~TeV (green)  and 10~TeV (blue). 
Assumed RG with the luminosity of 100$L_\odot$ and temperature of 3600~K.
    \label{fig:eloss}
  }
\end{figure}

  In Fig.~\ref{fig:eloss} we present numerical calculations of the energy loss of the electrons of different energies during their diffusion through the NSR.
  While in the early stages after the escape from the shock into the NSR the energy loss on RG dominates, it is not fast enough to cool down the electrons completely.
  As the electrons diffuse farther on, the CMBR losses start to dominate.
  This effect is particularly important at higher energies, where the Klein-Nishina effect further suppresses the energy losses on the RG radiation field.

The size of the electron NSR around Nova RS Oph in case of the Bohm diffusion and the IC energy losses on the CMBR is then estimated on,
\begin{eqnarray}
x_{\rm dif}/D = 0.67/B_{\rm \mu G}^{1/2} ~~~{\rm degree}.
%\label{eq2}
\end{eqnarray}
\noindent
   
      %which corresponds to the physical radius of $\sim 28.6/B_{\rm \mu G}^{1/2}$ pc
      %where $D = 2.45$ kpc is the distance to the Nova RS Oph.
      For the magnetic field of $3\mu G$, the size of the NSR around RS Oph is of the order of $\sim$0.4 degree, comparable to \textit{Fermi}-LAT angular resolution at a few GeV.
Note that this predicted size is of the order of the size of super-remnant optically detected in the case of 
M31N recurrent nova in M31 (see Darnley et al.~2019). The inner radius of the super-remnant shell in M31N is 52 pc which is not surprising since these recurrent novae clearly differ in basic parameters (e.g. the recurrence period of M31N is about ten times shorter which results in larger amount of energy provided by 
the nova into its super-remnant.

To validate those order of magnitude calculations we performed numerical tracking of electrons taking into account energy losses via IC on CMBR and RG radiation field (including also Klein-Nishina effect) as well as modification of the diffusion coefficient as the particle is losing energy.
Those calculations confirm that the diffusion process is limited by the CMBR rather than RG radiation field (however 5 GeV electrons still lose about 25\% of their energy in the latter process).
The obtained observed size of the $\gamma$-ray emission from those calculations is only slowly depending on the energy: $\sim0.24^\circ$ for $E_\gamma\lesssim0.1$~GeV and $\sim0.28^\circ$ for $E_\gamma\gtrsim1$~TeV.
Thus we conclude that the NSR around RS Oph is predicted to be a point source for the \textit{Fermi}-LAT and at most only moderately extended source for the Cherenkov telescopes.

At 200 GeV the angular resolution of CTA is $\sim 0.08^\circ$ \cite{mai17}.
Hence in this Bohm diffusion limiting case, the drop of the sensitivity of CTA for observations of such slightly extended emission will be by a factor of $\lesssim3$.
Therefore the extension of the emission might prevent the detection in the cases when the predicted emission is close to the point-like sensitivity limit of CTA.
However in most of the investigated cases with long activity time of the source and fast acceleration, the detection is still within the reach of CTA, even despite the possible extension.

\section{Conclusion}

We investigated the consequences of the hypothesis that electrons are accelerated in the shells ejected in novae for a long period after their explosions (see Bednarek~2022). We consider the models which cover a broad range of possible parameters for the acceleration process of the electrons within the shell, i.e. slow and fast acceleration and also short (or long) period injection of electrons after the nova explosion. As an example of the short period injection we consider the period of the order of a month after nova explosion (corresponding to the typical time scale of the GeV $\gamma$-ray emission observed by {\it Fermi}-LAT, and typically obtained in a number of nova acceleration models). The long period injection model assumes acceleration of electrons in the shell during the recurrence period of the nova RS Oph. Gamma-ray observations so far were unable to probe the acceleration over such long time scales. 
Two regions in the expending nova ejecta are also distinguished: (1) the equatorial region of the nova binary system in which the shell is effectively decelerated due to the entrainment of matter from the RG wind, and (2) the polar region in which the shell expands freely up to the moment of entrainment of the matter from the interstellar medium. As an example, electrons are assumed to be injected within the shell with the power law spectrum (spectral index -2, Bell~1978) and normalization to the $10\%$ of the kinetic energy of the nova shell.
Such an order of the energy conversion efficiency from the kinetic energy of the shell to particles is expected in the case of supernova remnants responsible for the cosmic ray content in the Galaxy 
(e.g. Schlickeiser~2002).
With such a general acceleration model we likely cover a broad range of possible specific acceleration scenarios which will in the future consider in more detail the evolution of the nova shell in long time after explosion. The $\gamma$-ray emission predicted above, for the range of models with considered by us boundary parameters, constrains possible emission expected from such future specific models.

Electrons lose energy on radiation processes in evolving environment of the shell.
Those, accelerated within the shell, can finally escape from the shells into the nova environment forming so called nova super-remnant. We concentrate on the case of the recurrent novae which, in such scenario, can periodically supply fresh relativistic electrons into the NSR. These electrons produce $\gamma$ rays via ICS of mainly optical radiation from the RG and the CMBR. The level of this $\gamma$-ray emission depends naturally on the activity period of the recurrent nova. 

We show that in the case of fast and efficient acceleration of the electrons and weak magnetization of the shell, the $\gamma$-ray emission from the NSR around the Nova RS Oph is expected to extend to TeV energy range.
The predicted emission is within the sensitivity of the CTA. 
In more optimistic cases (acceleration for the whole recurrence period and total activity period of at least $10^5$ yrs), there is a chance to detect the GeV $\gamma$-ray emission from NSR with \textit{Fermi}-LAT.
We note however, that the NSR is predicted to have the size of even $\sim 0.3$ degree
at GeV and TeV energies (if the electrons diffuse in the nova surrounding according to the Bohm diffusion prescription and the average magnetic field strength is $3\mu$G, i.e. comparable to the interstellar space).
So, NSR is expected to be moderately extended for $\gamma$-ray instruments.
However, the magnetic field in the NSR might be stronger, due to its supply with the incoming shells of nova.
This should essentially limit the size of the electron NSR.

In fact, direct correlated emission in the optical and $\gamma$-ray light curves of Nova V906 Car at early stage of the nova suggests that emission in this different energy ranges is produced by this same population of particles, likely relativistic electrons producing broad low energy emission in the synchrotron process and the $\gamma$-ray emission in the IC process (Aydi et al.~2020).

\section*{Acknowledgments}
We thank D. Green and R. Lopez-Coto for useful discussions and the Referee for many useful comments. 
This work is supported by the grant through the Polish National Research Centre No. 2019/33/B/ST9/01904.

\end{document}